\documentclass[10pt,journal,compsoc]{IEEEtran}
\usepackage{amsmath,amssymb,amsfonts}
\usepackage{algorithmic}
\usepackage[ruled,vlined]{algorithm2e}
\usepackage{graphicx}
\usepackage{textcomp}
\usepackage{xcolor}
\usepackage{subfigure}
\usepackage{epstopdf}
\usepackage{colortbl}
\usepackage{multirow}
\usepackage{enumitem}

\ifCLASSOPTIONcompsoc
  \usepackage[nocompress]{cite}
\else
  \usepackage{cite}
\fi
\ifCLASSINFOpdf
\else
\fi
\begin{document}
%
\title{Misplaced Subsequences Repairing with Application to Multivariate Industrial Time Series Data
}
%
%
%
%

\author{Xiaoou Ding,
        Hongzhi Wang,
         Jiaxuan Su, and Chen Wang.
\IEEEcompsocitemizethanks{\IEEEcompsocthanksitem XO Ding, HZ Wang, and JX Su are with Harbin Institute of Technology, P.O.Box 750, Harbin, Heilongjiang, 150001, China.

C Wang is with National Engineering Laboratory for Big Data Software, EIRI, Tsinghua University, Beijing, China.

E-mail: dingxiaoou@stu.hit.edu.cn, wangzh@hit.edu.cn, itx351@126.com,  wang\_chen@tsinghua.edu.cn
}
}

%
%

\markboth{Journal of \LaTeX\ Class Files,~Vol.~X No.~X, X~20XX}%
{Ding \MakeLowercase{\textit{et al.}}: Misplaced Subsequences Repairing with Application to Multivariate Industrial Time Series Data}
%



\IEEEtitleabstractindextext{%
\begin{abstract}
Both the volume and the collection velocity of time series generated by monitoring sensors are increasing in the Internet of Things (IoT). Data management and analysis requires high quality and applicability of the IoT data. However, errors are prevalent in original time series data. Inconsistency in time series is a serious data quality problem existing widely in IoT. Such problem could be hardly solved by existing techniques. Motivated by this, we define an inconsistent subsequences problem in multivariate time series, and propose an integrity data repair approach to solve inconsistent problems. Our proposed repairing method consists of two parts: (1) we design effective anomaly detection method to discover latent inconsistent subsequences in the IoT time series; and (2) we develop repair algorithms to precisely locate the start and finish time of inconsistent intervals, and provide reliable repairing strategies. A thorough experiment on two real-life datasets verifies the superiority of our method compared to other practical approaches. Experimental results also show that our method captures and repairs inconsistency problems effectively in industrial time series in complex IIoT scenarios.
\end{abstract}

}

\maketitle

\IEEEdisplaynontitleabstractindextext

%
\IEEEpeerreviewmaketitle

\IEEEraisesectionheading{\section{Introduction}\label{sec:introduction}}

\begin{figure}[t]
\centering

\includegraphics[scale=0.2352]{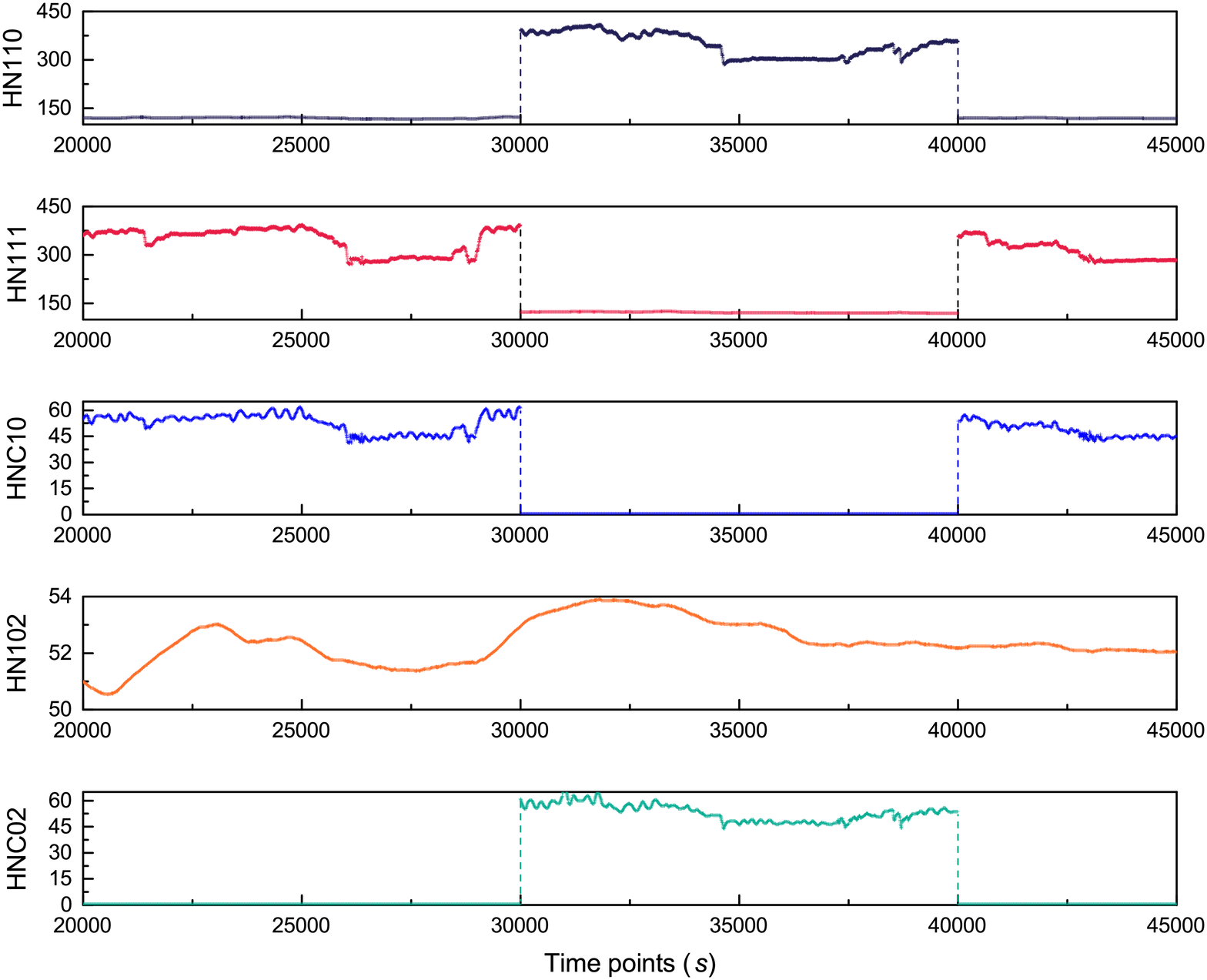}
\caption{An inconsistency example.}
\label{example1}
\end{figure}

\begin{figure}[t]
\centering
\includegraphics[scale=0.35]{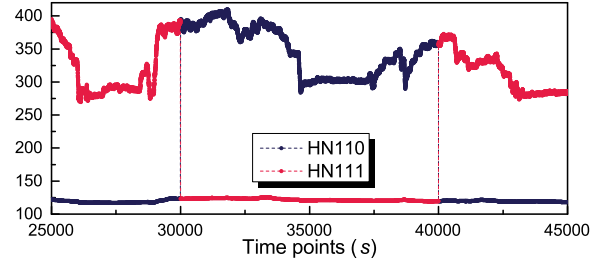}
\caption{Inconsistency between sequence HN110 and HN111.}
\label{example2}
\end{figure}

%
%
%
%
\IEEEPARstart{T}{his} widespread use of various monitoring sensors and the rapid performance improvement of sensing devices both give birth to data management and analysis in the Internet of Things (IoT). Time series data collected from sensor devices are one important data form in IoT. In data monitoring systems, data points are always collected together simultaneously from multiple dimensions, where each dimension (\emph{a.k.a}., attribute) corresponds to one sensor \cite{DBLP:journals/apin/DingLZ0L16}. Thus, the multi-dimension data from multiple sensors describe the status of a whole equipment together.
That is,
for a $M$-dimensional time series $\mathcal{S}$, the $m$-th sequence of $\mathcal{S}$ corresponds to the $m$-th dimension monitoring data.
\newline \indent As the high-quality IoT data is acknowledged to be the basic premise to achieve reliable information extraction and valuable knowledge discovery~\cite{DBLP:journals/pvldb/AbedjanCDFIOPST16}, the quality demand for time series data has grown stricter in various data application scenarios \cite{DBLP:conf/kdd/ToledanoCBT17,DBLP:series/synthesis/2014Gupta}. However, time series data are often dirty and contain quality problems, especially in industrial background. \cite{DBLP:journals/pvldb/DingWSLLG19} has proposed three kind of industrial time series data problems, namely missing values, inconsistent attribute values, and abnormal values or anomaly events. We have further investigate that misplaced subsequences in multivariate time series is one serious inconsistency problem during data quality management.
\newline \indent In real time series monitoring system \emph{e.g.,} Cyber-Physical Systems (CPS), some values in the $m$-th sequence may not correspond to the $m$-th dimension monitoring, due to the
unexpected troubles and signal interference during the undergoing working condition transition of the equipments. For example, clock errors may arise among sensors with different types. Transmission delay between sensors and the monitoring system also probably happens because of short-time network faults. In such cases,
a length of subsequence from $m_1$-th dimension may be recorded in the $m_2$-th dimension from a time point, and it will last for a time interval. This gives rise to an inconsistency problem during a certain working condition. We present a motivation example for an inconsistency instance below.
\newtheorem{example}{Example}
\begin{example}
Figure \ref{example1} shows a segment of sequences from five sensors of an equipment in the same time interval. We can find that subsequences in sensor HN110, HN111, HNC10, and HNC02\footnote{For privacy concern, we have made data desensitization of the name and the ID of sensors.} present abnormal sequence patterns, and they are possibly recorded incorrectly in the current sequence
in time interval $[30000s,40000s]$. Figure \ref{example2} partially enlarges the inconsistent subsequences existing in HN110 and HN111 in $[30000s,40000s]$. The fact is that a length of subsequence of HN110 is falsely recorded in HN111, while subsequence of HN111 is placed in HN110.
\end{example}

\textcolor[rgb]{0.00,0.07,1.00}{The aforesaid misplaced subsequences problem in time series data under industry scenarios bring at least two kinds of challenges.}
\begin{itemize}
  \item The misplaced subsequences problem belongs to \emph{continuous errors} \cite{DBLP:journals/access/WangW20}, other than happens in a single point. It is necessary to find out when the misplaced subsequences problem arises and how long it will last. As the time series data is collected continuously and densely, it is not easy to precisely compute the start and end time point of such inconsistency intervals. Method to identify these interval bounds must be sensitive and reliable enough for capturing the tendency of unexpected changes timeously.
  \item The pattern of misplaced subsequence errors is complicated. The data monitoring system often suffers different sensor failures or system errors, it is uncertain that how many attributes are involved in one misplaced error. As the number of attributes (sensors) of an equipment is not small, the increasing data amount and the attribute number add to the difficulty of both error detection and repair process.
\end{itemize}
Though data quality demand in time series is increasing, the research on repairing multi-dimensional inconsistent subsequences is not adequate. For time series data cleaning study,  outlier detection (or called anomaly detection, error identification, \emph{etc}) techniques have been developed for various application \cite{DBLP:series/synthesis/2014Gupta}. However, few methods are proposed for the data repairing tasks. A recent survey paper \cite{DBLP:journals/access/WangW20} has reviewed kinds of time series error cleaning methods. Most studies focus on the data cleaning of single point errors, and pay less attention to continuous errors in multivariate time series. For the existing data inconsistency repairing study, most techniques are mainly designed for relational data, and do not apply to inconsistency problems in time series.
\newline\indent  As the inconsistency in multivariate time series have just uncovered recently in time series management systems, especially in industry field, effective  solutions are still in high demand in abnormal patterns identification and inconsistent subsequence pairs repairing in multi-dimensional time series \cite{DBLP:journals/apin/DingLZ0L16}. 
Motivated by this, we address the problem of repairing inconsistent subsequences in multivariate time series under the industrial applications in this paper.
We summarize our \textbf{contributions} as follows:
\newline \indent (1) We extend  the misplaced 
formalize a serious inconsistency problem in Industrial Internet of Things (IIoT) data management,
\emph{i.e.}, inconsistent subsequences repairing in multivariate time series, according to real IIoT scenarios.
\newline \indent (2) We devise an integrated method to detect inconsistent time intervals in data collected by data monitoring systems, and correspond the inconsistent subsequences in each interval to the correct dimensions. Considering the real challenges in industrial data management, our method has the following accomplishments.
\begin{itemize}
  \item \textbf{Effectiveness in complex industrial scenarios}. 
      We design algorithms to detect inconsistent subsequences accurately from multi-dimensional time series under complex situations. Our method can well identifies and repairs hybrid inconsistent subsequences. (see Fig. \ref{frame} in Sec. \ref{sec3})
      \item \textbf{Less negative cumulative effect in sequence behavior modelling}. During the abnormal sequence behaviors process, our method distinguishes real inconsistent data from normal sequences with a well-designed sequence behavior model (see Algorithm \ref{alg2}). The proposed detection phase guarantees that the anomaly part will not effect the performance of the following detection. Moreover, our method always identifies inconsistent time intervals with both the start and the end points (we called them \emph{bounds} below). It guarantees the reliability of the solutions under industrial data repairing scenarios, because it will not modify those normal sequences by mistake.
  \item \textbf{Fault-tolerance repairing approach}. We propose a method to obtain repairing solutions from the evaluation of the candidate repair schemas (see Sec. \ref{sec4.2} and \ref{sec4.3}). In this step, our algorithms reconsider
      all the possible falsely processed schemas carefully with necessary modification (\emph{e.g.,} union or replace), and then provide high-quality repair solutions for true inconsistent time intervals.
\end{itemize}
\indent\indent (3) We conduct a thorough experiment on two real-life datasets from large-scale IIoT monitoring systems over 5 consecutive months. Experimental results on real-life data demonstrate the effectiveness of our method. Comparison experiments show that the proposed repairing strategies significantly improve both the accurate and the efficiency of the inconsistency repairing.
\newline \indent \textbf{Organization}.
The rest of the paper is organized as follows:
We introduce the related work in Sec. \ref{newsec2}, and discuss the basic definitions and the overview of our approach in Sec. \ref{sec2}. Sec. \ref{sec3} introduces inconsistent intervals detection approach and candidate repair schemas computation. Sec. \ref{sec4} discusses the evaluation on candidate repairing schemas and the determination of repairing results. Experimental study is reported in Sec. \ref{sec5}, and we draw our conclusion in Sec. \ref{sec6}.
\begin{figure*}[t]
\centering
\includegraphics[scale=0.3]{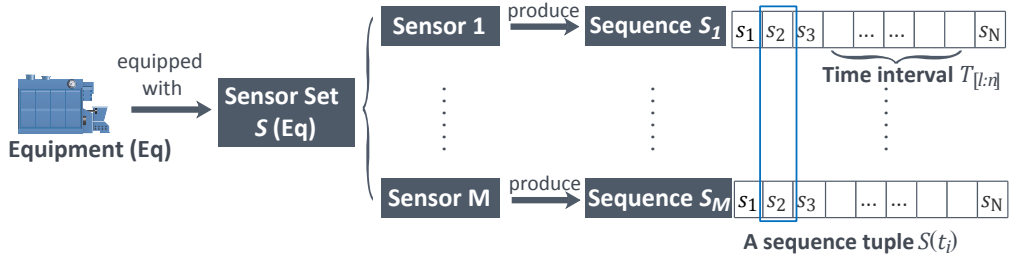}
 \caption{Multivariate IIoT time series.}
  \label{data}
  \end{figure*}
\section{Problem Overview}
\label{sec2}
We first define the inconsistent subsequences repairing problem in Sec.~\ref{sec2.1}, and introduce our solution framework of the proposed problem in Sec. \ref{sec2.2}.
\subsection{Preliminaries}
\label{sec2.1}
An equipment is normally regarded as the minimum independent unit of monitoring time series in IoT data management systems. According to \cite{DBLP:journals/apin/DingLZ0L16}, we outline the basic concepts in our problem with Fig. \ref{data}.
Each equipment has a sensor set, denoted by $\mathcal{S}_{\mathsf{Eq}}=\{S_1,...,S_M\}$, where each sensor $S_m (m \in [1,M])$ generates one time series along the time axis, and all sensors generate time series simultaneously. These time series are collected from the corresponding equipment and monitored by IoT data management systems. We define the sequence from a sensor in Definition~\ref{DS}, and the sensor set of an equipment is regarded as a multivariate time series, as shown in Definition \ref{MTS}. \newtheorem{definition}{Definition}
\begin{definition}
\label{DS}
(\textbf{Sequence}).
$S=\langle s_1,...,s_N\rangle$ is a sequence on sensor $S$, where $N=|S|$ is the length of $S$, \emph{i.e.}, the total number of elements in $S$. $s_n= \langle x_n,t_n\rangle, (n \in[1,N])$, where $x_n$ is a real-valued number with a time point $t_n$, and for $\forall n,k \in [1,N]$, it has $(n <k) \Leftrightarrow  (t_n <t_k)$.
\end{definition}
\begin{definition}
\label{MTS}
(\textbf{Multivariate time series}). Let $\mathsf{Eq}$ be an equipment sensor group. $\mathcal{S}_\mathsf{Eq}=\{S_1,...,S_M\} \in \mathbb{R}^{N\times M}$ is a $M$-dimensional time series, where $M$
is the total number of equipment sensors, \emph{i.e}., the number of dimensions.
\end{definition}
In this paper, we focus on one of the continuous errors, \emph{i.e.,} inconsistent subsequences in time series. According to Definition \ref{Subseq}, a subsequence $S_{[l,n]}$ corresponds to a time interval $T_{[l:n]}$ (Definition \ref{inter}). It is obvious that for a $M$-dimensional time series $\mathcal{S}$, $T_{[l:n]}$ provides $M$ subsequences from their corresponding sequences. All these subsequences share a common length, \emph{i.e.}, $T_{[l:n]}$'s length.
\begin{definition}
\label{Subseq}
(\textbf{Subsequence}).
A subsequence $S_{[l,n]} = \langle s_l, ... , s_{n} \rangle$, $(1\leq l\leq n <N)$ is a continuous subset of sequence $S$, which begins from the element $s_l$ and ends in $s_n$.
\end{definition}
\begin{definition}
\label{seqtuple}
(\textbf{Sequence tuple}).
A sequence tuple in a $M$-dimensional $\mathcal{S}$ is
the set of all data points at time $t_i$, denoted by $\mathcal{S}({t_i})=\langle s_{i1},s_{i2},...,s_{iM}\rangle$, \emph{i.e.}, the $i$-th row of $\mathcal{S}$.
\end{definition}
\begin{definition}
\label{inter}
(\textbf{Time interval}). Let $T=\{t_1,...,t_n\}$ be the set of time points of time series $\mathcal{S}_{\mathsf{Eq}}$,  $T_{[l:n]}$ is a time interval in $T$ which begins from time point $t_l$ and ends at $t_{n}$.
\end{definition}
In industrial data acquisition systems, the $M$-dimensional sequences of equipment $\mathsf{Eq}$ have a definite acquisition order, and they are recorded into a sequence $S_1,S_2,...,S_M$ correspondingly. As unexpected problems will cause inconsistency in some time intervals among multiple sensors, subsequences may be recorded into wrong dimensions during a period of time.
That is, the $M$ sequences are not orderly recorded into $S_1,S_2,...,S_M$ in time interval $T_{[l:n]}$. {On the basis of
a practical observation and investigation,} inconsistency in time intervals presents different patterns. Here, we apply the permutation structure \cite{Lyndon1977Combinatorial, DBLP:phd/ethos/McKay70} to describe the inconsistency pattern in Definition \ref{zhihuan}.
\begin{definition}
\label{zhihuan}
(\textbf{Permutation pattern}).
Given a time interval $T_{[l:n]}$ and the set of sequences with inconsistency problems \emph{i.e.,} $\mathcal{S}_{\textsc{inc}} =\{S_1,...,S_m\}, m\in[2,M)$,
an one-one mapping of $\mathcal{S}_{\textsc{inc}}$ to itself is regarded as a permutation of $\mathcal{S}_{\textsc{inc}}$, denoted by $\sigma: \mathcal{S}_{\textsc{inc}}\rightarrow \mathcal{S}_{\textsc{inc}}$, having
\begin{equation}\label{sub}
  \sigma  =\left (
  \begin{array}{c}
    S_i  \\
    S^{\sigma}_i
  \end{array}
   \right )=\left (
  \begin{array}{cccc}
    S_1 & S_2 & ... & S_m \\
    S^{\sigma}_1 & S^{\sigma}_2 & ... & S^{\sigma}_m
  \end{array}
   \right ) \nonumber
\end{equation}
\hfill $\square$
\end{definition}

\begin{example}
\label{e2}
Let HN110, HN111, HNC10, HN102, and HNC02 in Fig. \ref{example1} be sequence $S_1,S_2,S_3,S_4$, and $S_5$, respectively. According to Definition \ref{sub}, the inconsistency problem in time interval $[30000s,40000s]$ can be formalized as,
\begin{equation}
  \sigma  =\left (
  \begin{array}{cccc}
    S_1 & S_2 & S_3  & S_5 \\
    S_2 & S_1 & S_5  & S_3
  \end{array}
   \right ) \nonumber
\end{equation}
\hfill $\square$
\end{example}
Further, one permutation $\sigma$ may consists smaller structures, \emph{i.e.,} the permutation between $S_1$ and $S_2$. We introduce rotation pattern in Definition \ref{lunhuan}, which is indivisible and denoted as the unambiguous minimum repair unit in our method. Definition \ref{lunhuan} shows that a $m$-rotation $\sigma$ describes that each element $\alpha_i$ in $\mathcal{S}_{\textsc{inc}}$ is replaced by the next element $\alpha_{i+1}$, and the last element $\alpha_m$ is replaced by $\alpha_1$.

\begin{definition}
\label{lunhuan}
(\textbf{Rotation pattern}). $\sigma = \{ \alpha_1, \alpha_2, ... , \alpha_{m} \}$ is  a permutation pattern if having
 \begin{equation}\label{rotation}
  \sigma = \left (
  \begin{array}{cccc}
    \alpha_1 & \alpha_2 & ... & \alpha_{m} \\
    \alpha^{\sigma}_1 & \alpha^{\sigma}_2 & ... & \alpha^{\sigma}_{m}
  \end{array}
   \right ) =\left (
  \begin{array}{cccc}
    \alpha_1 & \alpha_2 & ... & \alpha_{m} \\
    \alpha_2 & \alpha_3 & ... & \alpha_1
  \end{array}  \right ) \nonumber
\end{equation}
Such $m$-rotation pattern is denoted by $\sigma(\alpha_1,\alpha_2,...,\alpha_{m})$, where $m$ is the order of $\sigma$, \emph{i.e.,} the number of elements in $\sigma$. \hfill $\square$
\end{definition}

Let $\sigma_1 (\alpha_1,\alpha_2,...,\alpha_{l})$ and  $\sigma_2 (\beta_1,\beta_2,...,\beta_{k})$ be two rotation patterns on $\mathcal{S}_{\textsc{inc}}$, $(l+k \leq m)$. According to the properties on \emph{permutation group} \cite{Lyndon1977Combinatorial},
$\sigma_1$ and $\sigma_2$ is disjoint if $\{\alpha_1,\alpha_2,...,\alpha_{l}\}$ and  $\{\beta_1,\beta_2,...,\beta_{k}\}$ differ from each other. Such disjoint $l$-rotation $\sigma_1$ and $k$-rotation $\sigma_2$ is indicated as
\begin{equation}\label{disjoint}
  \sigma_1 \cup \sigma_2 =(\alpha_1,\alpha_2,...,\alpha_{l})(\beta_1,\beta_2,...,\beta_{k})
\end{equation}
Now we apply rotation patterns to formalize an inconsistency instance in Definition~\ref{instance}.

\begin{definition}
\label{instance}
(\textbf{Inconsistency instance}). Let $\mathcal{S}_{\textsc{inc}}$ be the set of all inconsistent subsequences on $\mathcal{S}_{\mathsf{Eq}}$.
An inconsistency instance in a time interval $T_{[l:n]}$ is regarded as the union of a number of disjoint rotation patterns, describing the inconsistent patterns of $\mathcal{S}_{\textsc{inc}}$. It has
\begin{equation}
\varphi(I)=\sigma_1 \cup \sigma_2\cup\cdots\cup\sigma_k = \bigcup^{k}_{i=1} \sigma_i. \nonumber
\end{equation}
 where $\sum_{i=1}^k o_i \leq m (k\geq 1)$, and $o_i$ is the order of the $i$-th rotation pattern $\sigma_i$.
 $I=T_{[l:n]}$ is identified as an inconsistent time interval \emph{w.r.t.} $\varphi$.
 \hfill $\square$
\end{definition}

\begin{example}
\label{e3}
With the structure of rotation patterns, the inconsistency instance in Fig. \ref{example1} can be denoted as $\varphi(I) = \sigma_1 \cup \sigma_2=(S_1,S_2)(S_3,S_5)$. $I =T_{[30000s:40000s]}$ is an inconsistent time interval.
\hfill $\square$
\end{example}

It is worth noting that the properties of permutation group guarantee the uniqueness of the pattern of each inconsistency instance in interval $I$ as shown in Theorem \ref{theorem1}.
\newtheorem{theorem}{Theorem}
\begin{theorem}
\label{theorem1}
Given an inconsistent time interval $I$, the inconsistency instance $\varphi$ in $I$ has the unique form of the product of disjoint rotation patterns, which covers all inconsistent subsequences.  \hfill $\square$
\end{theorem}
Faced with inconsistency instances existing in time series from sensors, we aim to identify all inconsistent time intervals and repair all inconsistent instances correctly. We formalize the inconsistency repairing problems studied in this paper below, which consists of two tasks: the inconsistency detection problem and the inconsistency repair problem, respectively.
\begin{figure*}[t]
\centering
\includegraphics[scale=0.21]{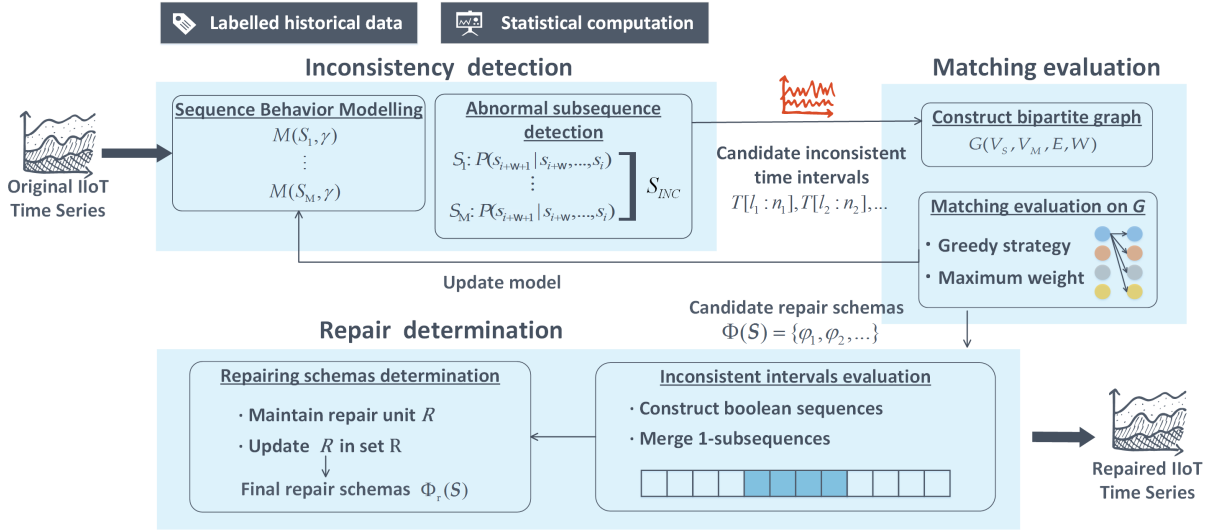}
\caption{Method framework overview}
\label{frame}
\end{figure*}
\newtheorem{problem}{Problem}
\begin{problem}
Given a $N$-length $M$-dimensional time series $\mathcal{S}$, the \textbf{inconsistency detection problem} on $\mathcal{S}$ is to find all $K$ inconsistent time intervals, denoted by $\mathcal{I}=\{I_1,....,I_K\}$, which satisfies
\newline \indent (1) $\forall I_i \in \mathcal{I},$ $I_i$ is the maximal interval covers one inconsistency instance $\varphi$; and
\newline \indent (2) $\forall$ $i,j \in[1,K]$ and $i\neq j$, $I_i$ and $I_j$ are two independent time intervals, \emph{i.e.}, $I_i \cap I_j = \emptyset$.
\end{problem}
\begin{problem}
The \textbf{inconsistency repair problem} on $\mathcal{S}$ is to compute the repair pattern of each inconsistent interval $I_i$ by identifying the inconsistency instance in $I_i$, which satisfies $\forall I_i \in \mathcal{I}$, $\varphi(I_i)$ covers all inconsistent subsequences denoted by the rotation patterns in $I_i$.
\end{problem}
\newtheorem{proposition}{Proposition}

\subsection{Method Framework}
\label{sec2.2}
Figure \ref{frame} illustrates the framework of our proposed solution, which consists of three phases: inconsistency detection, matching evaluation and repair determination.
\newline \indent The \emph{inconsistency detection} phase (see Sec. \ref{sec3.1}) is the first step in our method, where sequence behavior models are constructed to distinguish abnormal subsequences in each sensor from normal sequences. Parametric models are applied in our method according to priori knowledge or normality learning from historical IIoT data. We detect anomalies in each sequence with a sliding window, and an inconsistency instance is considered to exist in interval $T_{[l:n]}$ which contains a number of abnormal subsequences.
\textcolor[rgb]{0.00,0.07,1.00}{ Our inconsistency detection phase is open to most time series anomaly detection techniques, which will be presented in our experimental study.}
\newline \indent In the \emph{matching evaluation} phase (see Sec. \ref{sec3.2}), we compute possible repair schemas for all candidate inconsistent time intervals obtained from the previous step. In order to match inconsistent subsequences to their corresponding dimensions, we first construct a bipartite graph ${G}=(V_S,V_\mathcal{M},E,\mathcal{W})$. Each abnormal dimension in $I_i$ is presented as a source node in $V_S(G)$, and sequence models of the involved dimensions are treated as terminal nodes in $V_\mathcal{M}(G)$. We obtain repairing patterns with \emph{bipartite graph matching algorithms} on $G$.
\newline \indent \emph{Repair determination} is the most important phase in our proposed method (see Sec. \ref{sec4}). In this phase, we precisely locate each inconsistent time interval with start and end time points,
and provide accurate repair solutions. We propose algorithms to identify real inconsistent time intervals and provide final repair patterns. We apply the structure of disjoint set to obtain reliable repairing and effectively decreasing \emph{false positives} and \emph{false negatives}.
\newline \indent We summarize the notations frequently used in this paper in Table \ref{tab:1}.
\begin{table}[t]
\centering
\caption{List of frequent notations}
\label{tab:1}       
\begin{tabular}{|c|l|}
\hline
\cellcolor[rgb]{0,0,0}{\textcolor[rgb]{1.00,1.00,1.00}{\textbf{$\mathsf{Symbol}$}}} & \cellcolor[rgb]{0,0,0}{\textcolor[rgb]{1.00,1.00,1.00}{\textbf{$\mathsf{Description}$}}}  \\
\hline
$s$ & a data point in time series\\
$S$ &  the sequence generated by sensor $S$  \\
$S_{[l,n]}$ & subsequence beginning from $s_l$ and ending in $s_n$\\
\hline
$\mathcal{S}$ & (\emph{M}-dimensional) time series\\
$\mathcal{S}(t_i)$ & the sequence set of all data points at $t_i$ \\
$\mathcal{S}_{\textsc{INC}}$ & the set of inconsistent attribute\\
\hline
$t_i$ & a time point \\
$T_{[l:n]}$ & a time interval from $t_l$ to $t_n$.\\
$I$ & an inconsistent time interval\\
$\mathcal{I}$ & the set of inconsistent time interval on $\mathcal{S}$\\
\hline
$\sigma$ & a rotation repair\\
$\varphi(I)$ & the inconsistency instance on $I$.\\
$\Phi(\mathcal{S})$ & the set of candidate repair schemas for  $\mathcal{S}$ \\
$\Phi_\textrm{r}(\mathcal{S})$ & the set of final repair schemas for  $\mathcal{S}$ \\
\hline
$R$ & the repair unit of $\sigma$\\
\textbf{R} & the set of all repair units\\
\hline
 $B(\sigma)$ &  the boolean sequence for $\sigma$\\
 $B^1$ & a length of subsequence with all 1 elements \\
\hline
\end{tabular}
\end{table}

\section{Inconsistency Behavior Detection}
\label{sec3}
In this section, we first outline how we detect anomalies in sequences in Sec. \ref{sec3.1}, and discuss how to compute candidate repair patterns in Sec. \ref{sec3.2}.
\subsection{Abnormal subsequences modelling and detection}
\label{sec3.1}
Abnormal subsequences detection is of crucial importance in high-quality repairing solutions. The accurate identification of abnormal subsequences contributes to a high performance of repairing methods. We first construct behavior detection model for each sequence in $\mathcal{S}$. Sequence models are considered as priori knowledge provided by the equipment instructions, which can also be learned from historical data or labelled sample data. We use a 2-tuple function $\gamma$:$(\mathcal{F}, T)$ to describe time series modelling metrics. Here, $\mathcal{F}$ is the set of metric functions, including statistical variables (\emph{e.g.}, mean and variance), subsequence distance metrics and feature vectors of a sequence $S$ in the duration of a working condition. We formalize the sequence behavior model in Definition \ref{Model}.
\begin{definition}
\label{Model}
(\textbf{Sequence Behavior Model}).
Given a $M$-dimensional $\mathcal{S}=\{S_1,...,S_M\}$, the normal behavior of the $i$-th dimension sequence $S_i$ is modelled by $S_i \sim\mathcal{M}(S_i,\gamma)$, where $\gamma$:$(\mathcal{F}, T)$ is 2-tuple function for $S_i$.
\end{definition}

Accordingly, we present the basic assumption in Proposition \ref{proposition1} that any segment of a normal subsequence on $S$ should satisfies the model $\mathcal{M}(S,\gamma)$, and the conditional probability $p(s_{n+1} \models \mathcal{M} (S, \gamma )|s_n, ... , s_l ) \geq \theta_\mathcal{M}$ should be larger than a support threshold.
\begin{proposition}
\label{proposition1}
Given a {model support threshold} $\theta_\mathcal{M}$, if we have $S \sim \mathcal{M}(S,\gamma)$, then the following conditions are true:
\newline \indent (1) $\forall \, 1\leq l\leq n\leq N$, $S_{[l,n]} \models \mathcal{M}(S,\gamma)$, and
\newline \indent (2) $p (s_{n+1} \models \mathcal{M} (S, \gamma )|s_n, ... , s_l ) \geq \theta_\mathcal{M}$,
\newline where $S_{[l,n]}$ is a subsequence of $S$, and $s_i$ is a data point in $S_{[l,n]}, i\in[l,n+1]$. $\cdot\models\mathcal{M}$ denotes that the subsequence {corresponds with} model $\mathcal{M}$.\hfill $\square$
\end{proposition}
Now we are able to detect unexpected values in sequence and further discover latent inconsistent intervals according to Proposition \ref{proposition1}.
We detect abnormal data in each sensor sequence in $\mathcal{S}$ independently, where subsequence $S_{[l,n]}$ with $n-l+1$ continuous data point in $S$ is taken as a sliding window interval to determine whether there exists anomaly in the $(n+1)$-th window, \emph{i.e.}, $s_{n+1}$. Data $s_{n+1}$ is recognized abnormal when we detect $p(s_{n+1}\models \mathcal{M}(S,\gamma)|s_{n},...,s_{l})<\theta_\mathcal{M}$. Further, it is possible to be inconsistent when there exists some abnormal subsequences in a time interval $T_{[l:n]}$. And we will compute candidate repair results for $T_{[l:n]}$ with Algorithm 1 below.
\subsection{Candidate Repairing Schemas}
\label{sec3.2}
For the set $\mathcal{S}_{\textsc{inc}}=\{S_i,...,S_m\} (m\in[2,M])$ in interval $I$, $I$ is likely to contain $m$ inconsistent subsequences. Our task is to match each inconsistent subsequence to the correct sensor sequence. In general, we need to find $m$ one-one mapping between a subsequence and the correct sequence. We transform this matching problem into \emph{perfect matching} on a bipartite graph, and we construct the bipartite graph according to  Definition \ref{graph}.
\begin{definition}
\label{graph}
(\textbf{Bipartite graph construction}). Given the set $\mathcal{S}_{\textsc{inc}}$ on time $t$, and $\mathcal{M}(S_i)$ is the sequence model of the $i$-th element in $\mathcal{S}_{\textsc{inc}}$. $G=(V_{S},V_\mathcal{M},E,W)$ is a directed bipartite graph of $\mathcal{S}_{\textsc{inc}}$, where each element in $\mathcal{S}_{\textsc{inc}}$ is treated as a source node, \emph{i.e.}, $V_{S}=\{u_i|u_i\in \mathcal{S}_{\textsc{inc}}, i\in[1,m]\}$, and terminal nodes are the set of these $m$ sequence models,
denoted by $V_\mathcal{M}=\{\mathcal{M}(S_1),...,\mathcal{M}(S_m)\}$.
$e(u,v)$ describes a matching function from a subsequence to a sequence model $f: u\rightarrow \mathcal{M}(v)$, and the edge weight  $w(e)=p(u\models\mathcal{M}(v))$ represents the matching probability of $e(u,v)$.
\hfill $\square$
\end{definition}

After the construction of $G$, the repairing problem is transformed to {discovering optimization} matching on $G$. We consider two matching strategies to obtain candidate repair patterns. We first introduce an exact maximum weight matching solution. We then discuss a simple and fast greedy-based method, considering the balance between matching efficiency and effectiveness.
Intuitively, an inconsistent subsequence $S_{i.[l,n]}$ is recognized to only belong to one sequence $S_j$ with a quite high matching probability $p(S_{i.[l,n]}\models \mathcal{M}(S_j))$.
\newline \indent \underline{\emph{Maximum weight matching}}.
Considering our repairing problem, we need to find a \emph{maximum matching} \cite{DBLP:books/daglib/0023376} on $G$, which has $m$ one-one mapping between $V_S(G)$ and $V_\mathcal{M}(G)$. That is, to compute high-quality matching results with both maximum weights and maximum matching on $G$. We introduce the \emph{maximum cost maximum flow} (MCMF) algorithm \cite{DBLP:books/daglib/0023376} to compute the matching patterns. Accordingly, we add a global source node $\mathsf{s}$ and terminal node $\mathsf{t}$ to $G$. $\mathsf{s}$ points to all 0-in-degree nodes $V_S(G)$, and $\mathsf{t}$ is connected by all 0-out-degree nodes $V_\mathcal{M}(G)$. Clearly, an edge weight $w(u,v)$ represents the cost of a flow $u\rightarrow v$, \emph{i.e.,} the matching from $s_i$ to $\mathcal{M}(S_j)$. We obtain candidate repair patterns $\varphi$ by discovering a maximum matching on $G$ as follows.
\begin{equation}
\label{max}
\begin{split}
\max &\sum_{(u,v)\in E} cost(u,v)\cdot flow(u,v)\\
s.t. & \sum_{(u,v)\in E} flow(u,v)-\sum_{(u,v)\in E} flow(v,u) = f(u), \\
& 0\leq flow(u,v)\leq 1.
\end{split}
\end{equation}
where a feasible flow satisfies $\sum f(u)=0$, and $cost(u,v)= p(u\models \mathcal{M}(v,\gamma))$.
\newline \indent Note that the MCMF algorithm finds the maximum matching prior to maximum sum of weights, we can always achieve an one-one mapping from $V_S(G)$ to  $V_\mathcal{M}(G)$.
\newline \indent \underline{\emph{Greedy-based matching}}.
Intuitively, one inconsistency subsequence $S_{i.[l,n]}$ is recognized to only belong to one sequence $S_j$ with a quite high matching probability. In this case,
we design a heuristic greedy-based matching approach to achieve a fast matching on graph. When we make a match on $G$, we iteratively select  $u_i\rightarrow v_j$ which has the maximum edge weight computed by Equation (\ref{greedy}), and add this match $\varphi_{\mathsf{current}}$ to the result set. We then temporarily delete $u_i$ and $v_j$ from $G$. The matching process terminates until all nodes in $V_S$ have been matched to $V_\mathcal{M}$.
\begin{equation}\label{greedy}
  \varphi_{\mathsf{current}} =  \mathop{\arg\max}_{i, j\in [1,m]} w(u_i, v_j)=\mathop{\arg\max}_{s_i, s_j\in A} p(s_{i}\models \mathcal{M}(S_j,\gamma))
\end{equation}
\begin{algorithm}[t]
\label{alg2}
\caption{Compute Candidate Repairing Schemas}
\LinesNumbered 
\KwIn{a $N$-length $M$-dimensional time series $\mathcal{S}$, models for sequences in $\mathcal{S}$: $\mathcal{M}(\mathcal{S},\gamma)$, model support threshold  $\theta_{\mathcal{M}}$, a size number threshold $\epsilon$}
\KwOut{a set of candidate repair schemas on $\mathcal{S}$: $\Phi(\mathcal{S})$}
\ForEach{$t_i \in T$}
{initialize $\mathsf{cand}(\mathcal{S}(t_i)) \leftarrow  \mathcal{S}(t_i)$ and $A\leftarrow$ []\;
\For{$j$ from 1 to $M$}
{
\If{$p(s_{ij}\models \mathcal{M}(S_j,\gamma))< \theta_{\mathcal{M}}$}{$A\leftarrow  A\cup  \{j\};$}
\If{Size$(A)$ is no smaller than 2}{
initialize a matrix $\textbf{A}= |A|\times|A|$\;
\For{$n$ from 1 to $|A|$}{
\For{$m$ from 1 to $|A|$}{
$\textbf{A}_{nm}\leftarrow p(s_{iA_m}\models \mathcal{M}(S_{A_n}))$;}
}
construct $G$ according to $\textbf{A}_{nm}$\;
$\varphi \leftarrow$ matching result on $G$\;
\If{{Size$(\varphi) \leq \epsilon$}}
{$\varphi_i\leftarrow$ accept $\varphi$ as a candidate matching schema of $\mathcal{S}(t_i)$\;
repair $\mathsf{cand}(\mathcal{S}(t_i))$ with $\varphi_i$\;
}
\Else{{return $\mathcal{S}(t_i)$ and $ \varphi$ for artificial process}\;}}}
\ForEach{{$s_{j}' \in \mathsf{cand}(\mathcal{S}(t_i))$}}{merge $s_{j}'$ to $\mathcal{M}(S_j,\gamma)$ and update $\gamma$\;
move the sliding window to $[s_{j}',s_{j-1},....,s_{j-\mathsf{w}+1}]$;
}
}
\Return $\Phi(\mathcal{S})\leftarrow \{\varphi_i|i\in[1,N]\}$\;
\end{algorithm}

Algorithm~\ref{alg2} shows the process of computing candidate repairing schemas, which mainly consists of three phases: detecting abnormal behaviors (Lines 3-5), matching inconsistent subsequences (Lines 6-15) and updating sequence models with repaired data (Lines 18-20).

We first initialize a candidate repair set $\mathsf{cand}(\mathcal{S}(t_i))$ for each sequence tuple $\mathcal{S}(t_i)$, and maintain an array $A$ recording inconsistent data values at time point $t$. For the anomaly detection phase, the $\mathsf{w}$-length set of sequence tuples $\{\mathcal{S}(t_{i-1}),...,\mathcal{S}(t_{i-\mathsf{w}})\}$ ahead of $\mathcal{S}(t_i)$ serves as the sliding window to detect the model behavior of $\mathcal{S}(t_i)$. As each inconsistency instance happens in multiple sequences at the same time, we begin our detection \emph{simultaneously} and \emph{independently} in all $M$ dimensions within  sequence tuple $\mathcal{S}(t_i)$. For each sensor dimension $S_j$, we compute the probability of the current data corresponding to this dimension according to the modelling analysis $\mathcal{M}(S_j,\gamma)$ in Definition \ref{Model}. We insert the unexpected data point $s_{ij}$ into the inconsistency list $A$ if the probability $p(s_{ij}\models \mathcal{M}(S_j,\gamma))$ is smaller than a given threshold $\theta_{\mathcal{M}}$. It reveals  that $s_{ij}$ is unexpected to be recorded in sequence $S_j$.
\newline\indent With the discovered abnormal data points, $\mathcal{S}(t_i)$ is possible to be inconsistent if there exists several abnormal data points in $\mathcal{S}(t_i)$, \emph{i.e.,} the size of set $A=\{s_{i1},...,s_{in}\}$ is larger than 1. In this case, we construct a square matrix $\mathbf{A}$ for $\mathcal{S}(t_i)$, where the number of rows (resp. columns) is equal to the number of elements in $A$ (Lines 8-10). For each inconsistent data point $s_i$ in $A$, we compute the probability of modelling $s_i$ to each sequence involved in $A$ and {record $p(s_{im}\models \mathcal{M}(S_{n}))$ to the corresponding element in $\mathbf{A}$.}
\newline\indent Since we aim to match inconsistent subsequences to correct dimensions with the maximum likelihood,
we construct $G$ according to the matching probability matrix $\mathbf{A}_{nm}$, and obtain a match result $\varphi$ between $V_S$ and $V_\mathcal{M}$ (Lines 11-12). We check the total number of elements in $\varphi$, \emph{i.e.}, $\emph{Size}(\varphi)$, and accept $\varphi$ as a candidate repair schema $\varphi_i$ for a sequence tuple $\mathcal{S}(t_i)$ if $\emph{Size}(\varphi)$ is no larger than a given threshold $\epsilon$.  When $\emph{Size}(\varphi)> \epsilon$, we terminate this matching schema and return $\mathcal{S}(t_i)$ to human. This is because greater number of $\emph{Size}(\varphi)$ possibly reveals some complex anomalies or faults from the equipment sensor group, rather than inconsistency problems. Data will be returned to monitoring system engineers. We expect to obtain reliable decision and repairing result under such complex unexpected cases with knowledge engineering methods from domain experts.
\newline \indent After we have accepted $\varphi_i$ and matched inconsistent data to correct sequences, we insert the repaired data values to the sequence model and update $\gamma$, in order to improve accuracy of anomaly detection on following sequence data with the correct parameters. This step guarantees that statistical metric values will not be affected by the abnormal data values. We then successively move the sliding window and process next $\mathcal{S}(t_{i+1})$ in $\mathcal{S}$  (Lines 18-20) with the above steps.
Algorithm \ref{alg2} finishes after we process all time points and obtain the candidate repair schema $\Phi(\mathcal{S})$ for the \emph{N}-length time series $\mathcal{S}$.
\section{Repairing Solution}
\label{sec4}
When we aim to achieve an accurate and reliable repair of all inconsistency instances, it requires
an effective determination of inconsistency intervals. Accordingly, we design a step of determining final repair patterns with two main tasks: \emph{i}) to locate both the start and end timestamps of an inconsistent interval, and \emph{ii}) to repair each inconsistent interval with reliable schemas. However, both tasks are challenging to be completely solved in Algorithm \ref{alg2}. The reasons are discussed as follows.


For the former task, we need to further evaluate and merge the candidate schemas on sequence tuples to accurately detect the location of inconsistency intervals. Note again that an inconsistency instance always lasts a duration, rather than happen in several discrete time points. Thus, a reliable repair solution of an inconsistent interval should cover all data points within the interval. Since that sequence behavior modelling is analyzed by sliding window $(s_{n+1}|s_n,...,s_l)$ in Algorithm \ref{alg2}, it cannot always provide a uniform and accurate repair schemas for one inconsistency interval for the foregoing reasons.

For the latter, continuous high-quality repair schemas are difficult to be obtained from matching pattern evaluation in inconsistent industrial data. On the one hand, abnormal behaviors are not easily be to detected and distinguished from normal data for a sequence tuple in industrial time series. If the algorithms fail to precisely find the set $\mathcal{S}_{\textsc{inc}}$ for sequence tuple $\mathcal{S}(t_i)$ (see lines 4-5 in Algorithm \ref{alg2}), we will consequently obtain wrong matching results from the incorrect set $\mathcal{S}_{\textsc{inc}}$.
On the other hand, bipartite graph matching algorithms may run into \emph{partial} mismatch in some sequence tuples.  Both cases add to the number of either false positives or false negatives, and further result in a poor repair of $\mathcal{S}$.
\newline \indent To achieve an accurate and robust inconsistency repairing result, we propose a repairing schemas determination algorithm ($\mathsf{DRS}$) to precisely locate inconsistency intervals and further effectively repair inconsistent subsequences.
We first indroduce $\mathsf{DRS}$ algorithm in Sec. \ref{sec4.2}, and then discuss how to determine inconsistent intervals both effectively and efficiently in Sec. \ref{sec4.3}.
\subsection{Determining Repairing Schemas}
\label{sec4.2}
As discussed in Sec. \ref{sec2.1}, an inconsistency instance contains no less than one disjoint rotation patterns. Each rotation pattern is both indivisible and unambiguous. In order to effectively detect inconsistent intervals and identify inconsistency patterns, we introduce a repair unit in Definition \ref{triple} which serves as the minimum process unit in our method.

\begin{definition}
\label{triple}
\textbf{A repair unit} is a triple of a rotation pattern $\sigma$, denoted by $R$:$[\sigma,\textbf{T}, \emph{Size}(\textbf{T})]$. $\textbf{T}=\{T_{[l_1:n_1]},T_{[l_2:n_2]},...\}$ is the set of time intervals which are detected to be repaired by $\sigma$, and \emph{Size}$(\textbf{T})$ is the total number of time points in set $\textbf{T}$.
\end{definition}
Accordingly, a candidate repair schema $\varphi$ can be divided into several disjoint rotation patterns \emph{i.e.}, $\sigma_1,\sigma_2,\cdot\cdot\cdot$. We create and maintain the repair unit $R$ of each rotation $\sigma$ to evaluate all candidate inconsistent time intervals and determine the final repair schemas on them. The repair schemas determination is outlined in Algorithm \ref{algDRS}, which consists of two steps: \emph{i}) updating the set of repair units according to all divided $\sigma$s (Lines 2-7) and \emph{ii}) repairing subsequences in all inconsistent intervals (Lines 10-15).
\newline \indent We first enumerate each candidate repair schema $\varphi_i$ from $\Phi(\mathcal{S})$, and divide $\varphi_i$ into rotation patterns according to Theorem \ref{theorem1}. We create a repair unit for each $\sigma$ and record the location of such time intervals that are computed to be repaired by $\sigma$ as well as the total lengths of these intervals $\emph{Size}(\textbf{T})$ from Algorithm \ref{alg2} (Lines 4-6). After we obtain all repair units $\mathbf{R}$ from $\Phi(\mathcal{S})$, we sort all repair units in descending order according to $\emph{Size}(\textbf{T})$ and abandon those units which are used in candidate inconsistent intervals with a low frequency (Lines 8-9).
\newline\indent With the selected repair units set $\textbf{R}'$ in line 9, we further determine the accurate location of inconsistent intervals which contains rotation pattern $\sigma$ by processing algorithm $\mathsf{IIE}(\Phi(\mathcal{S}),\sigma)$ (see Algorithm 3 below).
After that, we enumerate each independent interval $I$ from $I(\sigma)$, in which we combine all accepted rotation patterns into a final integrated repair schema $\varphi_r$ and make the final repair of $\mathcal{S}$. After all repair units are processed,  Algorithm \ref{algDRS} finishes and returns high-quality time series $\mathcal{S}_\textrm{r}$ along with all repairing schemas $\Phi_\textrm{r}(\mathcal{S})$. 
\begin{algorithm}[t]
\label{algDRS}
\caption{Determining Repair Schemas}
\LinesNumbered 
\KwIn{the candidate $\Phi(\mathcal{S})$, {schema applied  interval length threshold:} $\mathsf{len1}$}
\KwOut{$\Phi_\textrm{r}(\mathcal{S})$}
Initialize $\mathbf{R}\leftarrow$ []\;
\ForEach {$\varphi_i \in \Phi(\mathcal{S})$}{
\ForEach {$\sigma \in \varphi_i$}{
\If{$\sigma$ does not exist in $\mathbf{R}$}{
create a triple $R: [\sigma, \textbf{T}, \emph{Size}(\textbf{T})]$ for $\sigma$\;
$\mathbf{R} \leftarrow \mathbf{R} \cup R$\;
}
update $\emph{Size}(\textbf{T})$ in $\mathbf{R}$\;
}}
Sort all repair triples in $\mathbf{R}$ in descending order of $\emph{Size}(\textbf{T})$\;
{
$\mathbf{R}'\leftarrow $ select repair triples by \emph{Size}$(\textbf{T})\geq \mathsf{len1}$}\;
\ForEach {$R \in \mathbf{R}'$}{
$I(\sigma)\leftarrow$
$\mathsf{IIE}(\Phi(\mathcal{S}),\sigma)$;
// see Algorithm \ref{alg32}.\\
\ForEach{$I \in I(\sigma)$}{
  repair $I$ with $\sigma$ and update $\mathcal{S}_\textrm{r}$ with repaired $I$\;
  $\varphi_\textrm{r}(I) \leftarrow \varphi_\textrm{r}(I) \cup \sigma$\;
  $\Phi_\textrm{r}(\mathcal{S}) \leftarrow \Phi_\textrm{r}(\mathcal{S}) \cup R$\;
}}
\Return $\mathcal{S}_\textrm{r}$ and $\Phi_\textrm{r}(\mathcal{S})$;
\end{algorithm}
\subsection{Inconsistency Intervals Evaluation}
\label{sec4.3}
We now introduce how to detect the accurate location of inconsistency intervals.
As discussed above,
we enumerate to evaluate each repair unit of a rotation pattern $\sigma$ (Line 10 in Algorithm \ref{algDRS}). During the process, we need to label which time intervals contains inconsistency pattern $\sigma$ and which does not.
We propose a boolean sequence $B$ for rotation pattern $\sigma$ in Definition \ref{bool}, which can assist to identify and extract inconsistent intervals.

\begin{definition}
\label{bool} (\textbf{Boolean sequence of $\sigma$}).
Given a $N$-length $M$-dimensional $\mathcal{S}$, $B(\sigma)=\langle b_1,...,b_N\rangle$ is a boolean sequence \emph{w.r.t.} rotation pattern $\sigma$, where $b_i \, (i\in [1,N])$ is a binary value assigned according to $\sigma$ as follows,
\begin{equation}\label{boolean}
  b_i = \left \{
\begin{array}{cl}
  1, & \sigma \, \textrm{exists in} \, t_i
  \\
  0, & \textrm{otherwise}.\\
\end{array}
\right.
\end{equation}
where $t_i$ is the $i$-th time point of $\mathcal{S}$. $B(\sigma)$ has the same length  with $\mathcal{S}$, \emph{i.e.,} $\emph{Size}(B(\sigma))=\emph{Size}(\mathcal{S})= N$. \hfill $\square$
\end{definition}
From the above, element $b_i = 1$ in $B(\sigma)$ represents that rotation $\sigma$ is adopted at time point $t_i$, while 0 means $\sigma$ is not adopted at $t_i$ or no inconsistency happens in $t_i$.  An intuitive observation is, either 0s or 1s in $B(\sigma)$ trends to continuously appear and make up a time interval. We denote a subsequence only consisting 0 (resp. 1) as 0-sequence block \emph{a.k.a} $B^{0}$ (resp. 1-sequence block $B^{1}$).  Accordingly, $B(\sigma)$ covers alternating appearance of $B^{0}$ and $B^{1}$, denoted by $B(\sigma) = \{ ..., B^{0}_{i}, B^{1}_{i+1}, B^{0}_{i+2} , ...\}$.
\newline \indent It is easy to discover subsequence $B^1$ when the element 1 continuously  and uninterrupted lasts for a number of time points in $B(\sigma)$. However, things are not simple when element 1s and 0s are {intertwined} in a period of time. It can be concluded that there exists falsely recorded 1s or 0s, for the reason that the occurrence of inconsistency instances always continues for a time duration, rather than happen in a quite short period of time. Such cases include two false patterns:
 \emph{i}) $b_i$ is a \emph{false positive} (FP) where the normal pattern is falsely detected to be inconsistent and repaired by $\sigma$, or \emph{ii}) $b_i$ is a \emph{false negative} (FN) where the inconsistency are falsely identified to be normal.
\newline \indent Faced with both problems, we consider a metric $\tau$ to measure whether a $B^1$ should be merged into its neighbor $B^0$ or not. In order to identify all real $B^1$s, \emph{i.e.}, the real inconsistent intervals with rotation $\sigma$, we evaluate all $B^0$s and $B^1$s with Equation (\ref{01}).
\begin{equation}
\label{01}
 \tau = \frac{|B_{i+1}|}{|B_{i}|+|B_{i+2}|},
\end{equation}
where $B_{i}, B_{i+1}, B_{i+2}$ are three continuous subsequences in $B(\sigma)$, and $|B_{i}|$ is the length of $B_{i}$.
\begin{algorithm}[t]
\label{alg32}
\caption{Inconsistent Intervals Evaluation}
\LinesNumbered 
\KwIn{the candidate $\Phi(\mathcal{S})$,  $\sigma$, $\theta_\tau$, {minimum inconsistency interval length threshold:} $\mathsf{len2}$ }
\KwOut{$I(\sigma)$: the inconsistent intervals set repaired by $\sigma$}
Initialize boolean sequence $B=\langle b_1,...,b_N\rangle$ and
a disjoint set $\mathcal{D}$ with $d_k.root \leftarrow k$\;
\ForEach {$\varphi \in \Phi(\mathcal{S})$}{
\If{$ \sigma(\alpha_1,...,\alpha_n)\in \varphi$ and $\forall \alpha_i \in \sigma$ has not been repaired}{$b_k\leftarrow 1$;}
\Else{$b_k\leftarrow 0$\;}
}
\ForEach{$b_j \in B (j\geq 2)$}{
\If{$b_j = b_{j-1}$}{$\mathcal{D}.\textsc{Union}(b_j,b_{j-1})$;}
}
$\mathbf{B}\leftarrow \{d_x| d_x \textrm{ is {the current independent element} in } \mathcal{D}\}$\;
\ForEach{$B^*_i \in \mathbf{B}$}{
\For{$ \mathsf{Bool}  \in \{1,0\}$}
{
\If{the boolean value of ${B^*_i}$ equals to $\mathsf{Bool}$ \textrm{and}
$\tau > \theta_\tau $}{
label each element in $B^*_{i+1}$ with $\mathsf{Bool}$\;
$\mathcal{D}.\textsc{Union}(B^*_i,B^*_{i+1})$, $\mathcal{D}.\textsc{Union}(B^*_{i+1},B^*_{i+2})$\;
}
}
}
\ForEach{$d_x \in \mathcal{D}$}{
\If{$b_{d_x} = 1$ and Size$(d_x)<\mathsf{len2}$}{
$I(\sigma)\leftarrow  I(\sigma) \cup d_x.T$;}
}
\Return {$I(\sigma)$};
\end{algorithm}
\newline \indent Now we present inconsistent intervals evaluation process in Algorithm \ref{alg32}. We evaluate each $\varphi$ with the involved rotation patterns in $\varphi$, and generate the boolean sequence of each $\sigma$ according to Definition \ref{bool} (Lines 3-6). 
We then begin to detect inconsistent intervals by determining all real $B^1$s with the start and end time points from $B(\sigma)$. For efficiency optimization, we use \emph{Disjoint Set} structure \cite{DBLP:books/daglib/0023376} to gather a subsequence $B^1$ (resp. $B^0$) with elements 1 (resp. $0$) in lines 7-9,
and further, we decide whether a $B^0$ should be merged into its neighbour $B^1$ or vice versa (Lines 11-15).
\newline \indent After $B(\sigma)$ is kept as $\{ ..., B^{0}_{i}, B^{1}_{i+1}, B^{0}_{i+2} , ...\}$ with disjoint structure $\mathcal{D}$, we copy $B(\sigma)$ to a set $\mathbf{B}$ and make further modification on $\mathbf{B}$ to avoid breaking the original structure in $B(\sigma)$. In the loop lines 11-15, we enumerate each element $B^*_i$ from set \textbf{B}, and compute $\tau$ according to Equation (\ref{01}). We merge the current $B^*_i$ with its neighbor block with the union function $\mathcal{D}$ if $\tau$ is smaller than a given threshold $\theta_\tau$  (Lines 14-15). It illustrates that the size of $B^*_{i+1}$ is too small to support its boolean value here, and the real value of $B^*_{i+1}$ should be replaced by its neighbor blocks \emph{i.e.}, $B^*_i$ and $B^*_{i+2}$. After the whole union process, we enumerate the updated 1-sequence blocks. If the block length is no smaller than $\mathsf{len2}$, \emph{i.e.,} it has enough length to be identified as an inconsistency instance, this block will be inserted into the inconsistent intervals set $I(\sigma)$ (Lines 17-18). Algorithm \ref{alg32} finishes until all sequence blocks in $B(\sigma)$ have been processed.
\section{Experimental study}
\label{sec5}
We now evaluate the experimental study of the proposed methods. All experiments run on a computer with 3.40 GHz Core i7 CPU and 32GB RAM.
\subsection{Experimental Settings}
\label{sec5.1}
\textbf{Data source}. We conduct our experiments on real-life industrial equipment monitoring data collected from two large-scale power stations. Details are shown in Table \ref{summary}.
\newline \indent (1) \underline{\textsl{FPP-sys}} dataset
 describes five main components of one induced draft fan equipment with 48 attributes from a large-scale \underline{f}ossil-fuel \underline{p}ower \underline{p}lant. Data on more than 1050\emph{K} historical time points for  5 consecutive months are applied in our sequence behavior model. We report our experimental results of repairing inconsistency  on 50\emph{K} time points.
\newline \indent (2) \underline{\textsl{WPP-sys}} dataset has  150 attributes describing the working condition of fan-machine groups from a wind power plant. It collects data each 8 seconds, and 1620\emph{K} time points data has been used in modelling process and we detect and repair inconsistency in 75\emph{K} time points data in the experiments.
\newline\indent \textbf{Implementation}.  We have developed \emph{Cleanits}, a data \underline{clean}ing system for \underline{i}ndustrial \underline{t}ime \underline{s}eries in our previous work \cite{DBLP:journals/pvldb/DingWSLLG19}, where three IoT data cleaning and repairing functions are implemented under real industrial scenarios.
\newline \indent We implement all algorithms of the proposed method in this paper as named $\mathsf{ISR}$. Besides, we implement another three algorithms for comparative evaluation:
\begin{itemize}
       \item $\mathsf{G}$-$\mathsf{ISR}$ uses greedy-based algorithm in bipartite graph matching in Algorithm \ref{alg2}, with the other steps the same as $\mathsf{ISR}$;
       \item $\mathsf{CRS}$ only executes Algorithm \ref{alg2} with maximum weight matching and outputs $\Phi(\mathcal{S})$ and $\mathsf{cand}(\mathcal{S})$ as the final repair result;
       \item $\lambda$-$\mathsf{Block}$ blocks the $N$-length $\mathcal{S}$ into small-length intervals, and takes each interval data as a whole part in behavior modelling process. The following steps are the same as $\mathsf{ISR}$.  The appropriate length of blocked intervals is $\lambda\cdot\sqrt{N}, \lambda\in[1,10]$.
           We report the experimental result with $\lambda=1$ as the best performance of $\lambda$-$\mathsf{Block}$ method.
\end{itemize}
\begin{table}[t]
\setlength{\abovecaptionskip}{0.cm}
\setlength{\belowcaptionskip}{-0.cm}
\caption{Summary of datasets}
\label{summary}
\centering
\scalebox{1}{
\begin{tabular}{|c|c|c|c|}
\hline
\textbf{Dataset} & \#\textbf{Sensors} & \#\textbf{Modelling }  & \#\textbf{Detection}\\
\hline
\textsl{FPP-sys} & 48 & 1050\emph{K} for 5 months & 50\emph{K} for 5 days
\\
\textsl{WPP-sys} & 150 & 1620\emph{K} for 5 months & 75\emph{K} for 7 days
\\
\hline
\end{tabular}}
\end{table}
\indent\indent \textbf{Measure}.  Since that the solution of inconsistency repairing problems contains both detection and repairing tasks, we evaluate and report algorithm performance in detection phase and repairing phase independently.
We apply Precision ($\mathsf{P}$) and  Recall ($\mathsf{R}$) metrics to evaluate the performance of all comparison algorithms.
In detection phase, we evaluate how well the algorithm identifies inconsistency intervals with Equation (\ref{PRd}). $\mathsf{P}_\mathsf{d}$ measures the ratio between the number of inconsistent intervals correctly detected and the total number of intervals detected by algorithms.  $\mathsf{R}_\mathsf{d}$ is the ratio between the number of intervals correctly detected and the total number of all inconsistent intervals. In repairing phase, we report the repairing quality with Equation (\ref{PRr}). $\mathsf{P}_\mathsf{r}$ computes the ratio between the number of inconsistent intervals correctly repaired and the total number of inconsistent intervals correctly detected from the above detection phase. Similarly, $\mathsf{R}_\mathsf{r}$ is the ratio between the number of correct repairs and the number of all detected inconsistent intervals.

It is worth noting that we pay more attention to the identification quality of inconsistent time intervals in detection evaluation, while we focus on the repair quality of concrete inconsistency instances in repairing results.
\newline \indent
\begin{equation}\label{PRd}
\mathsf{P}_\mathsf{d}= \frac{\textrm{\#}{\textsl{correctDectection}}}{\textrm{\#}\textsl{Dectection}}, \mathsf{R}_\mathsf{d}= \frac{\textrm{\#}\textsl{correctDectection}}{\textrm{\#}\textsl{InconsistentIntervals}}.
\end{equation}
\begin{equation}\label{PRr}
\mathsf{P}_\mathsf{r}= \frac{\textrm{\#}\textsl{correctRepair}}{\textrm{\#}\textsl{correctDectection}}, \mathsf{R}_\mathsf{r}= \frac{\textrm{\#}\textsl{correctRepair}}{\textrm{\#}\textsl{Dectection}}.
\end{equation}
\subsection{Evaluation on Real Errors}
\label{sec6.2}
\textbf{General performance}.
\label{sec6.2.1}
Table \ref{twodata} shows the repair performance of algorithms  on the two datasets, with \#\textsl{Time points} = 45\emph{K} in \emph{FPP}-\emph{sys} and \#\textsl{Time points} = 60\emph{K} in \emph{WPP}-\emph{sys}. Experimental results for algorithm $\lambda$-$\mathsf{Block}$ with varying $\lambda$ show that the appropriate length of blocked intervals is $\lambda\cdot\sqrt{N}, \lambda\in[1,10]$. We show the experimental result with $\lambda=1$ below as the best performance of $\lambda$-$\mathsf{Block}$ method.
Table \ref{twodata} shows that our proposed $\mathsf{ISR}$ has the highest performance of $\mathsf{P}_{\mathsf{r}}$ and $\mathsf{R}_{\mathsf{r}}$ on the two datasets. The repair recall of $\mathsf{ISR}$ is slightly higher than the precision. $\mathsf{CRS}$ comes the second, and its repair performance is a little better than $\mathsf{G}$-$\mathsf{ISR}$. It demonstrates that the greedy-based matching applied in $\mathsf{G}$-$\mathsf{ISR}$ is not as reliable as the maximum weight matching in $\mathsf{ISR}$. For $\mathsf{CRS}$, without further evaluation on candidate repair schemas,  it fails to provide high-quality repair results as $\mathsf{ISR}$ does. It is not surprising that $\mathsf{Block}$ has low costs on both datasets. However, the repair quality of $\mathsf{Block}$ is poor with the lowest repair precision of these four algorithms.

\begin{figure}[t]
\centering
 \subfigure[$\mathsf{P}_\mathsf{d}$, \textsl{FPP-sys}]{
\includegraphics[width=1.652in]{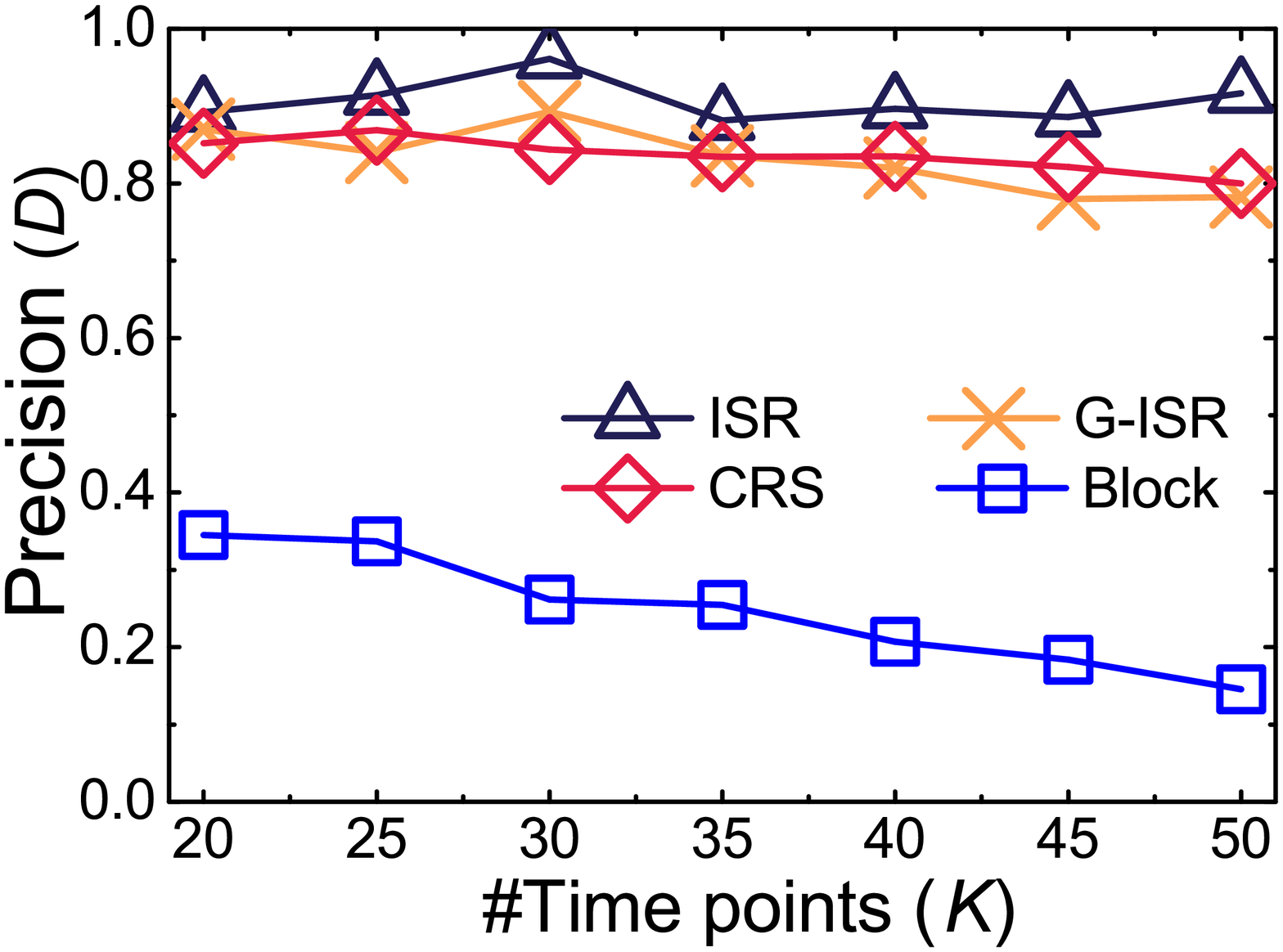}}
\subfigure[$\mathsf{R}_\mathsf{d}$, \textsl{FPP-sys}]{
\includegraphics[width=1.652in]{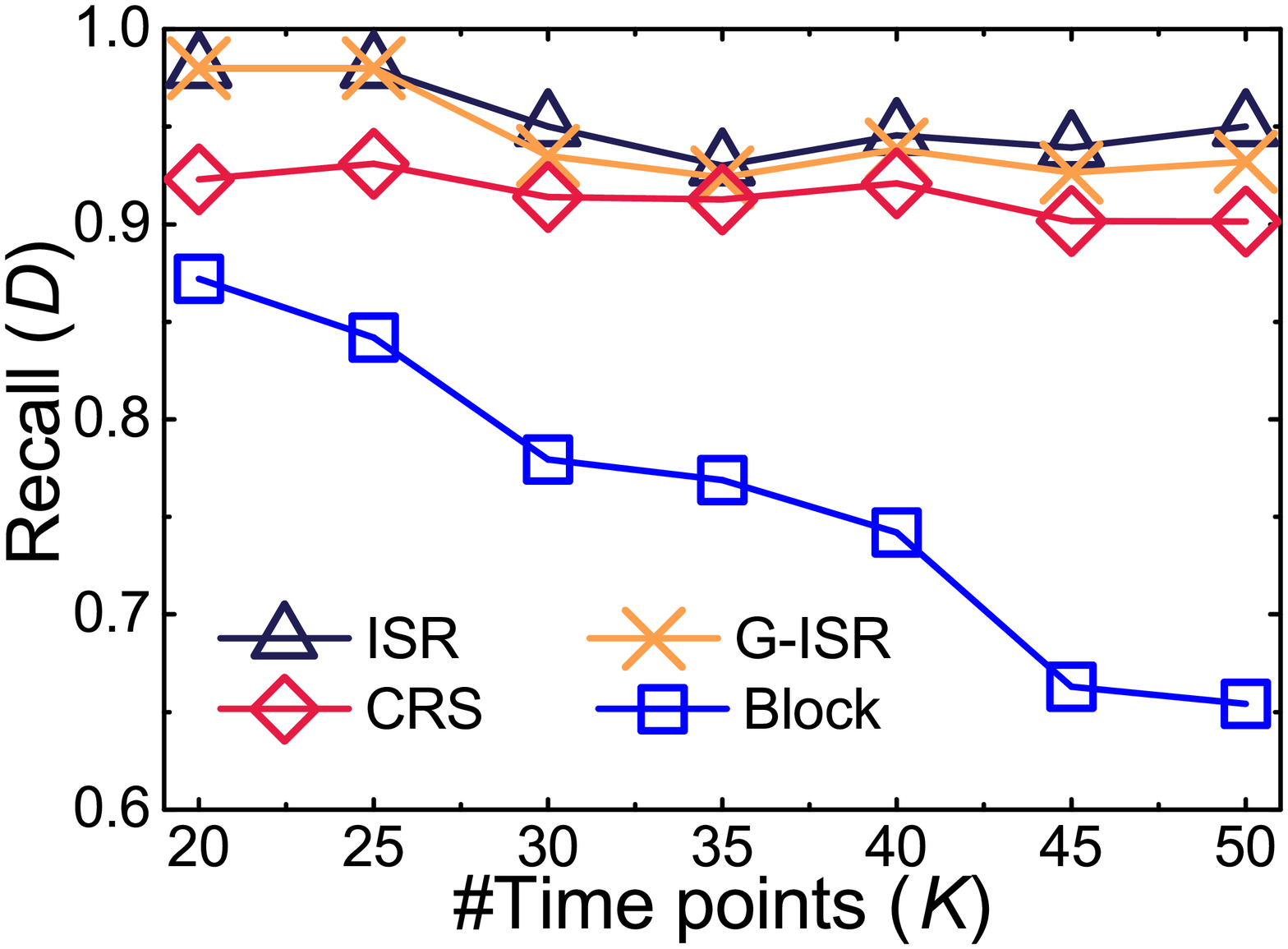}}
\caption{Inconsistency detection performance comparison vs. data volume}
\label{prd1}
\end{figure}

\begin{figure}[t]
\centering
 \subfigure[$\mathsf{P}_\mathsf{r}$, \textsl{FPP-sys}]{
\includegraphics[width=1.652in]{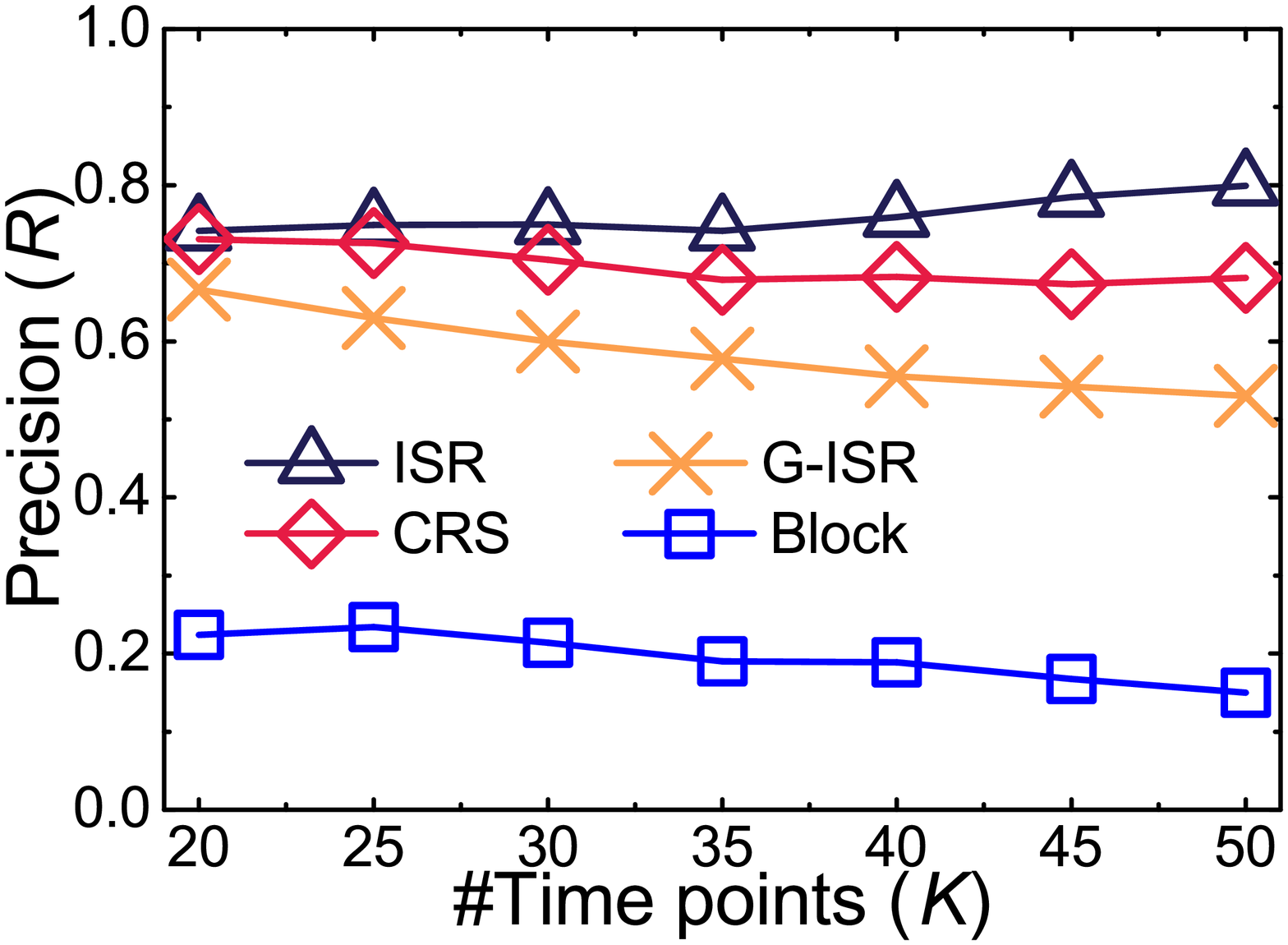}}
\subfigure[$\mathsf{R}_\mathsf{r}$, \textsl{FPP-sys}]{
\includegraphics[width=1.652in]{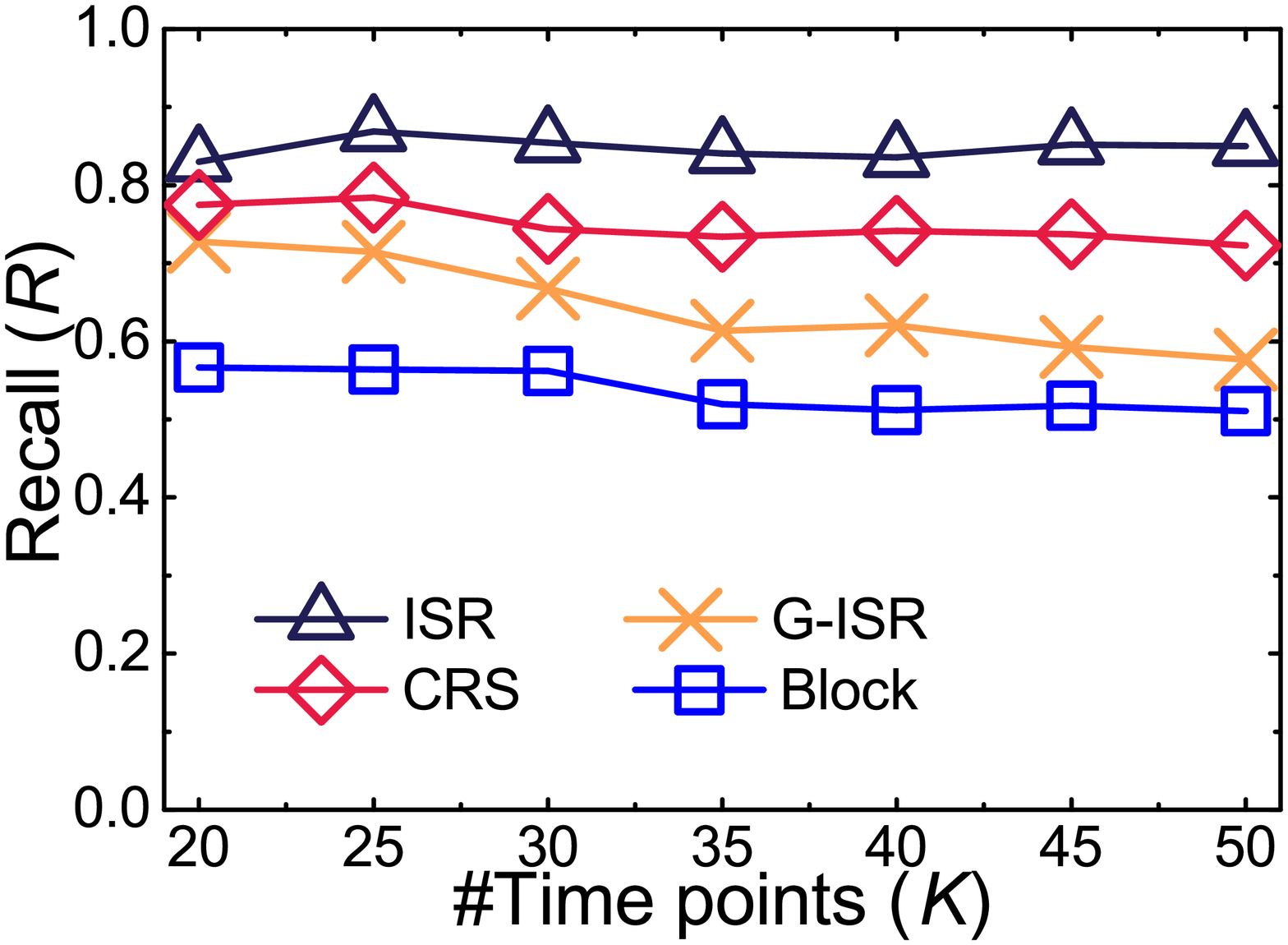}}
\caption{Inconsistency repairing performance comparison vs. data volume}
\label{prd2}
\end{figure}

We next report the detailed experimental results of all algorithms on the \textsl{FPP}-\textsl{sys} dataset with two important parameters: total data amount (\emph{i.e.}, \#\textsl{Time points}) and the maximum amount of inconsistent attributes (\emph{i.e.}, \#\textsl{Inconsistent Attr}).
\begin{table}[t]
\setlength{\abovecaptionskip}{0pt}
\setlength{\belowcaptionskip}{0pt}
\caption{Algorithms comparison on two datasets}
\label{twodata}
\centering
\scalebox{1}{
\begin{tabular}{|c|c|c|c|c|c|c|}
\hline
 & \multicolumn{3}{c|}{\textsl{FPP-sys} (45\emph{K})} & \multicolumn{3}{c|}{\textsl{WPP-sys} (60\emph{K})}
 \\ \cline{2-7}
 & $\mathsf{P}_\mathsf{r}$ & $\mathsf{R}_\mathsf{r}$ & Time(s) & $\mathsf{P}_\mathsf{r}$ & $\mathsf{R}_\mathsf{r}$ & Time(s) \\
\hline
$\mathsf{ISR}$ & \textbf{0.788} & \textbf{0.852} & 147.93 & \textbf{0.782} & \textbf{0.877} & 163.45\\
\hline
$\mathsf{G}$-$\mathsf{ISR}$ & 0.542 & 0.592 & 132.67 & 0.612 &  0.650& 138.56\\
\hline
$\mathsf{CRS}$ &  0.595 & 0.737 & 147.1 & 0.715 & 0.720 & 161.25\\
\hline
$\mathsf{Block}$ & 0.17 & 0.517 & \textbf{112.34} & 0.316 & 0.623 & \textbf{119.61}\\
\hline
\end{tabular}}
\end{table}
\newline \indent \textbf{Varying data amount}.
\label{sec6.2.2}
We report algorithm performance comparison on data volume varying form 20\emph{K} to 50\emph{K} in \textsl{FPP-sys} dataset with various inconsistent time intervals. On the condition that \#\textsl{Inconsistent Attr} = 12,
Figure \ref{prd1} and Figure \ref{prd2} show the performance on inconsistency detection and repairing, respectively.
\newline \indent Figure \ref{prd1} reveals that with the increasing data amount, the proposed $\mathsf{ISR}$ can always well detect inconsistent intervals, and  outperforms the other three methods on both $\mathsf{P}_\mathsf{d}$ and $\mathsf{R}_\mathsf{d}$. When \#\textsl{Time points} reaches 40\emph{K},  $\mathsf{ISR}$'s  $\mathsf{P}_\mathsf{d}$ maintains around 0.9, while $\mathsf{R}_\mathsf{d}$ keeps 0.92. It verifies that the proposed inconsistency detection method as well as the fault-tolerance strategy really contribute to a high-quality repairing of inconsistent intervals in industrial time series data.
\newline \indent Method $\mathsf{G}$-$\mathsf{ISR}$ and $\mathsf{CRS}$ come the second place on both $\mathsf{P}_\mathsf{d}$ and $\mathsf{R}_\mathsf{d}$. For $\mathsf{CRS}$, it outputs the repairing schemas from Algorithm \ref{alg2} as the final results without a further fault-tolerance strategy. This results in the poor performance of $\mathsf{CRS}$ compared with $\mathsf{ISR}$. Figure  \ref{prd1}(a) shows that $\mathsf{P}_\mathsf{d}$ of $\mathsf{CRS}$ never reach 0.9 and it has a downtrend with the increasing data volume. For $\mathsf{G}$-$\mathsf{ISR}$, the simple greedy-based matching approach does not make enough interval identification as the maximum weight matching does. It is because that $\mathsf{G}$-$\mathsf{ISR}$ sometimes computes incorrect candidate repair schemas, and consequently, the further inconsistent interval evaluation process fails to always provide reliable results. Method $\mathsf{Block}$ comes the least, and both metrics fell seriously with the growing data volume. It verifies that inconsistency problems cannot be detected well by such blocking method. As a crucial parameter affecting the quality of sequence behavior models,
the appropriate blocking length of intervals is challenged to be determined. In addition, some inconsistency instances with a small number of inconsistent attributes are difficult to be discovered by $\mathsf{Block}$.
\newline \indent  Figure \ref{prd2} shows inconsistency repairing performance with the same experimental condition. It shows that $\mathsf{ISR}$ has the best performance on both $\mathsf{P}_\mathsf{r}$ and $\mathsf{R}_\mathsf{r}$. Both metrics of $\mathsf{ISR}$ keep steadily with the increasing data amount, while $\mathsf{G}$-$\mathsf{ISR}$  shows a decline trend in either $\mathsf{P}_\mathsf{r}$ or $\mathsf{R}_\mathsf{r}$.  $\mathsf{CRS}$ always outperforms $\mathsf{G}$-$\mathsf{ISR}$, for the reason that $\mathsf{G}$-$\mathsf{ISR}$ trends to make more false matching on bipartite graphs with the increasing inconsistency instances in data. The stable metric values of both $\mathsf{ISR}$ and $\mathsf{CRS}$ show  that the proposed non-aftereffect sequence behavior modelling in Algorithm \ref{alg2} really helps to avoid incorrect anomaly detection results. Further, the performance difference between $\mathsf{ISR}$ and $\mathsf{CRS}$ highlights the necessary of the fault-tolerance repairing strategy proposed in Sec. \ref{sec4}. $\mathsf{ISR}$ really improves the repair effectiveness by evaluation of all candidate repair schemas.
\newline \indent \textbf{Varying inconsistent attributes}.
\label{sec6.2.2}
\begin{figure}[t]
\centering
 \subfigure[$\mathsf{P}_\mathsf{d}$, \textsl{FPP-sys}]{
\includegraphics[width=1.652in]{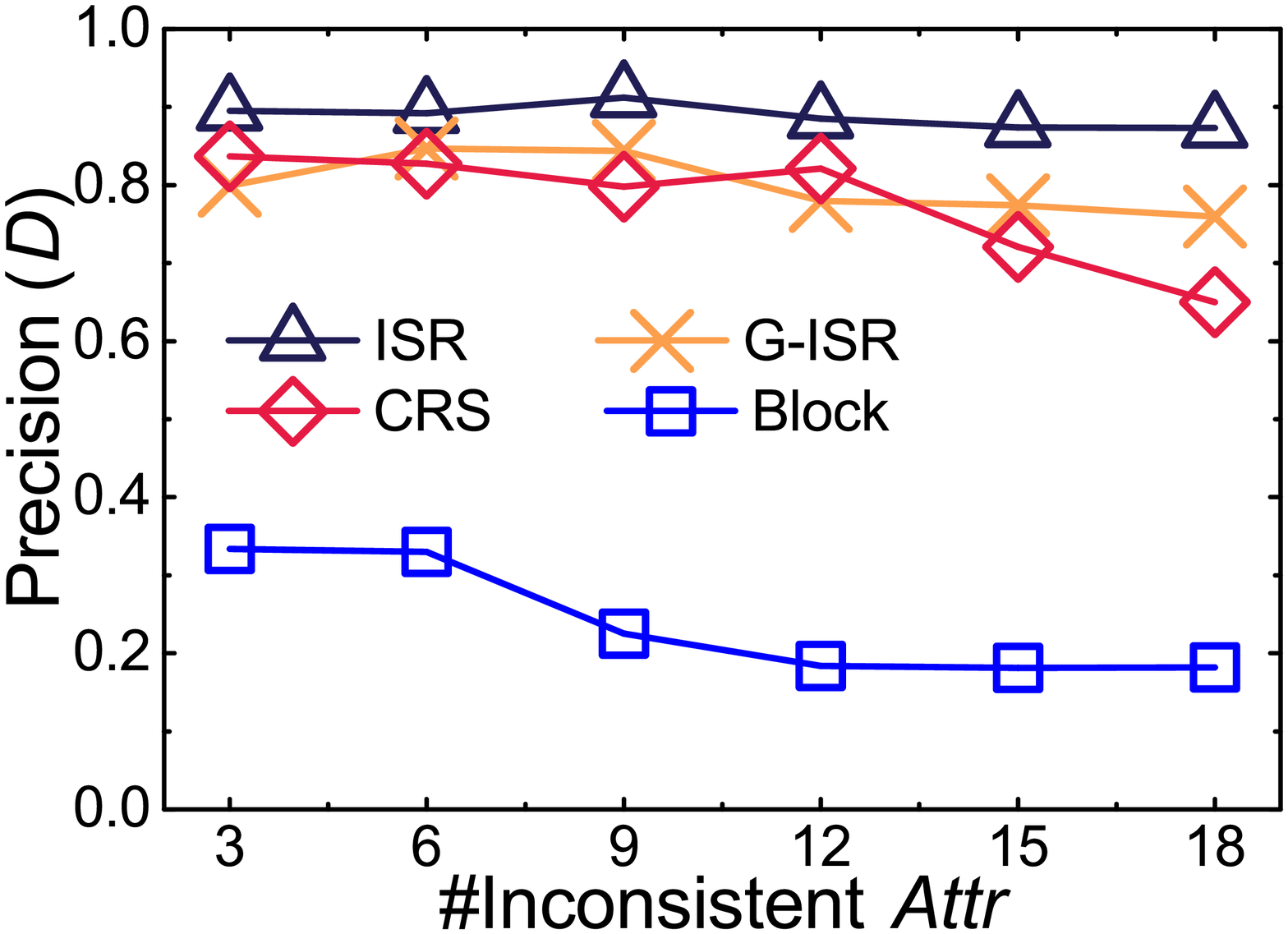}}
\subfigure[$\mathsf{R}_\mathsf{d}$, \textsl{FPP-sys}]{
\includegraphics[width=1.652in]{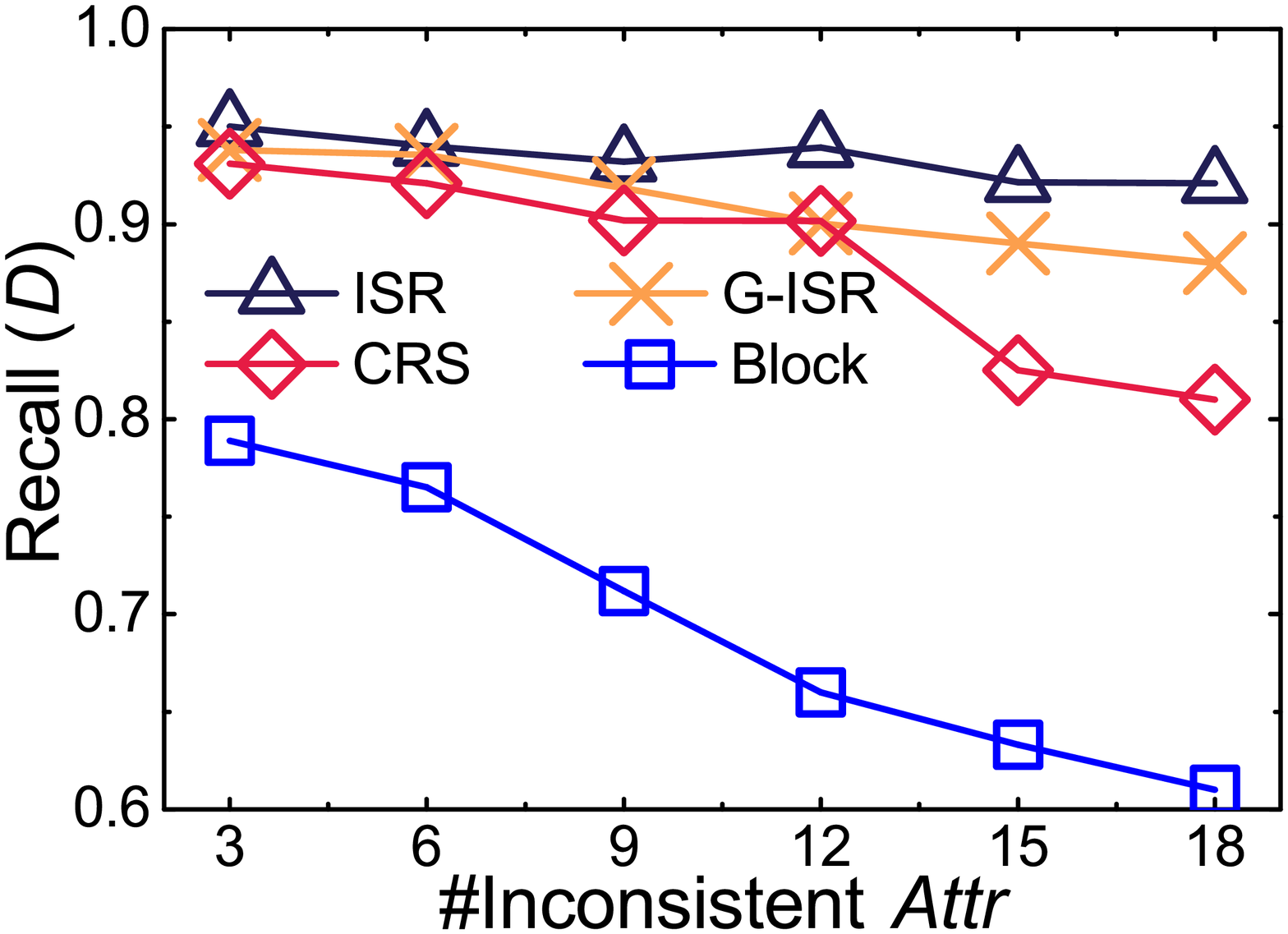}}
\caption{Inconsistency detection performance comparison vs. the number of inconsistent attributes}
\label{prr1}
\end{figure}
\begin{figure}[t]
\centering
 \subfigure[$\mathsf{P}_\mathsf{r}$, \textsl{FPP-sys}]{
\includegraphics[width=1.652in]{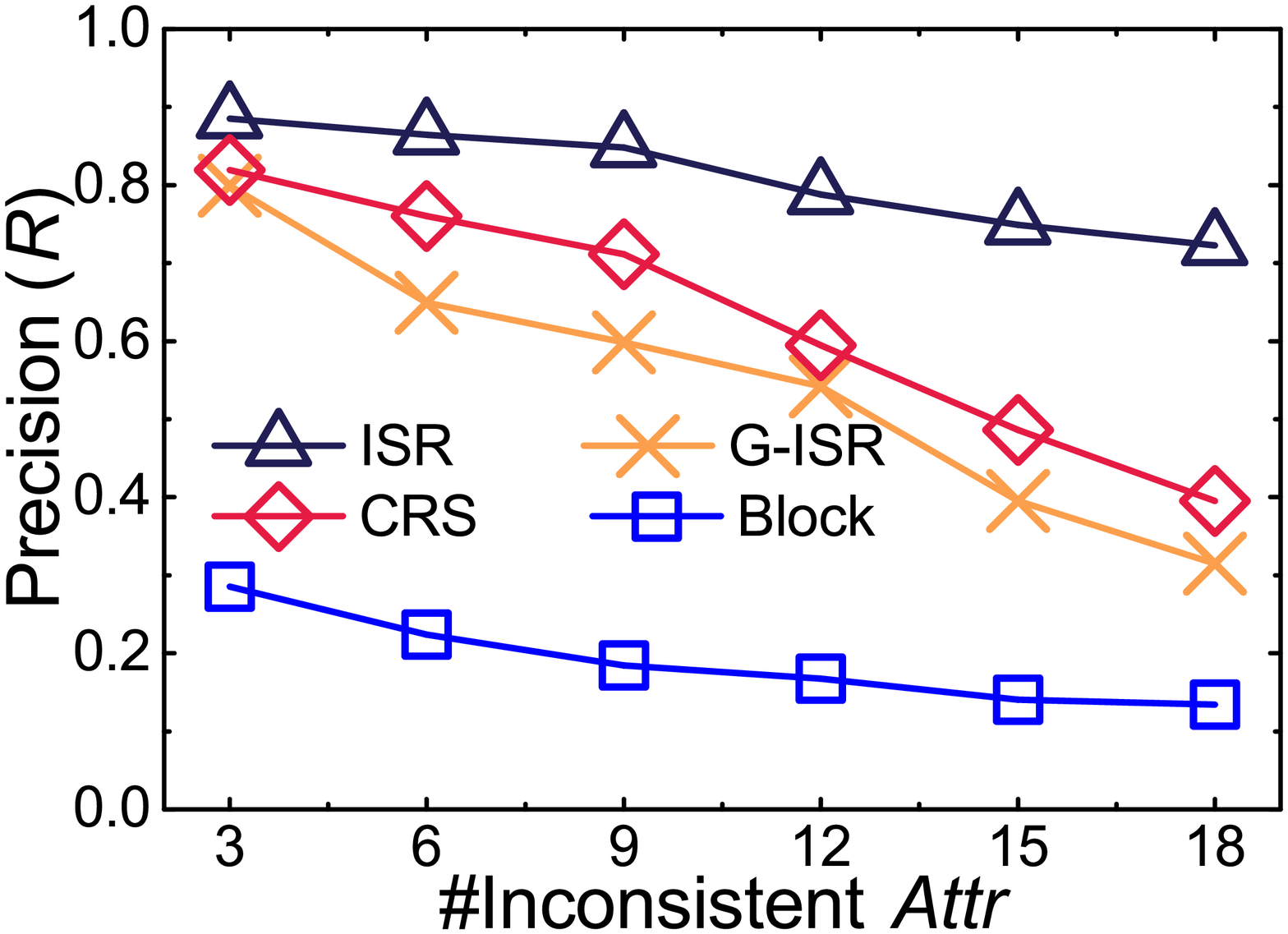}}
\subfigure[$\mathsf{R}_\mathsf{r}$, \textsl{FPP-sys}]{
\includegraphics[width=1.652in]{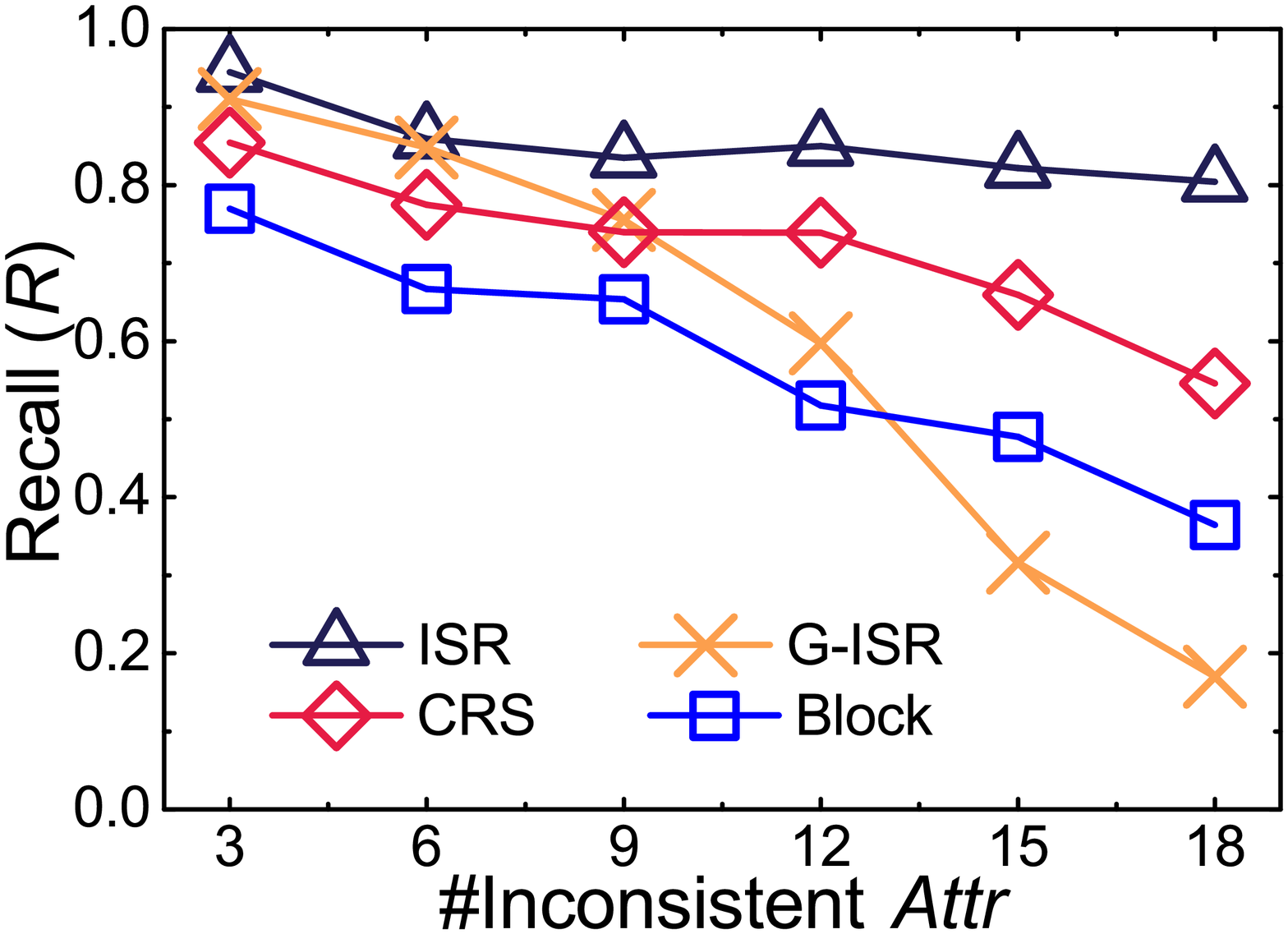}}
\caption{Inconsistency repairing performance comparison vs. the number of inconsistent attributes}
\label{prr2}
\end{figure}
Figure \ref{prr1} and Figure \ref{prr2} report the performance on the condition that \#\textsl{Time points} = 45\emph{K} in \textsl{FPP}-\textsl{sys} with \#\textsl{Inconsistent Attr} varying from 3 to 18.
\newline \indent Figure \ref{prr1}(a) shows that $\mathsf{ISR}$ has the highest $\mathsf{P}_\mathsf{d}$ and it keeps 0.87-0.91 against the increasing number of inconsistent attributes. $\mathsf{ISR}$'s repair recall only presents a slight drop when \#\textsl{Inconsistent Attr} is larger than 12.
$\mathsf{CRS}$ has a serious drop in the two metrics, which reflects the detection quality of $\mathsf{CRS}$ decreases when there exists inconsistency instances with more attributes.
In general, Figure \ref{prr1} confirms that our method is effectiveness in detecting inconsistency issues in monitoring industrial data under complex conditions.
\newline \indent Figure \ref{prr2} shows the repairing performance comparison among four methods. All methods suffer a drop in both $\mathsf{P}_\mathsf{r}$ and $\mathsf{R}_\mathsf{r}$ with the increasing number of inconsistent attributes.
Our $\mathsf{ISR}$ can still achieve $\mathsf{P}_\mathsf{r}> 0.78$ and $\mathsf{R}_\mathsf{r}> 0.85$ with the maximal inconsistency instances existing in 12 attributes simultaneously. The results verify that $\mathsf{ISR}$ are able to repair inconsistent instances effectively from low-quality data with multiple inconsistent attributes.
These metric values are acceptable and those false-repaired instances can be returned to artificial evaluation process as mentioned in Sec. \ref{sec3.2}.
\newline \indent Compared with $\mathsf{ISR}$,
both $\mathsf{G}$-$\mathsf{ISR}$ and $\mathsf{CRS}$ suffer a sharp drop with the increasing number of inconsistent attributes. And $\mathsf{G}$-$\mathsf{ISR}$ never performances better than $\mathsf{ISR}$. This illustrates that 1) the proposed $\mathsf{ISR}$
can detect inconsistent instances effectively from low-quality data with multiple inconsistent attributes, and 2) the greed-based matching approach is not as effective as the maximum weight approach, and it gets even worse in the cases that inconsistency happens in much more attributes.
\newline \indent \textbf{Efficiency}.
\label{sec6.2.3}
\begin{figure}[t]
\centering
 \subfigure[Execution time vs. data volume]{
\includegraphics[width=1.652in]{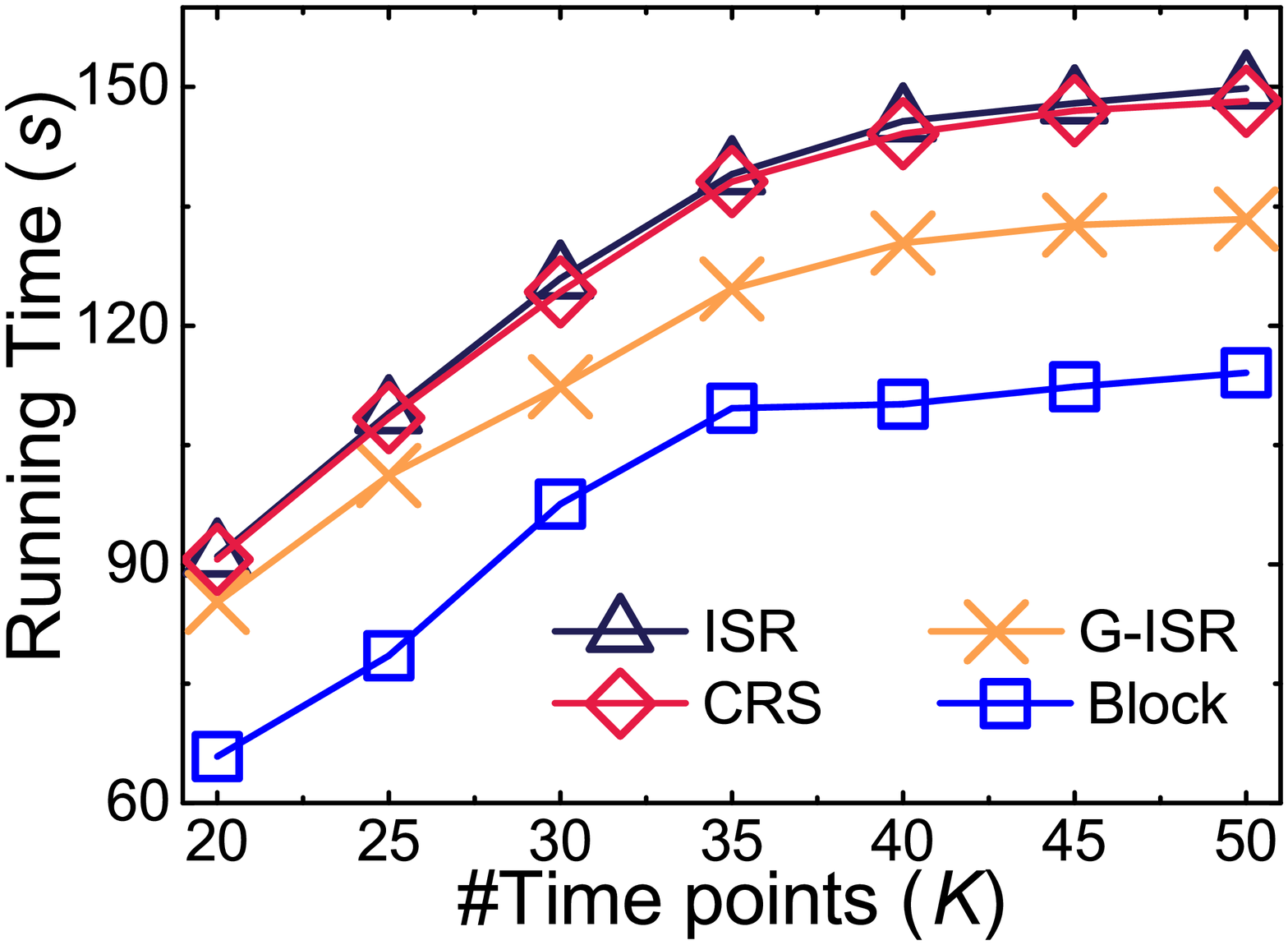}}
\subfigure[Execution time vs. inconsistent attribute number]{
\includegraphics[width=1.652in]{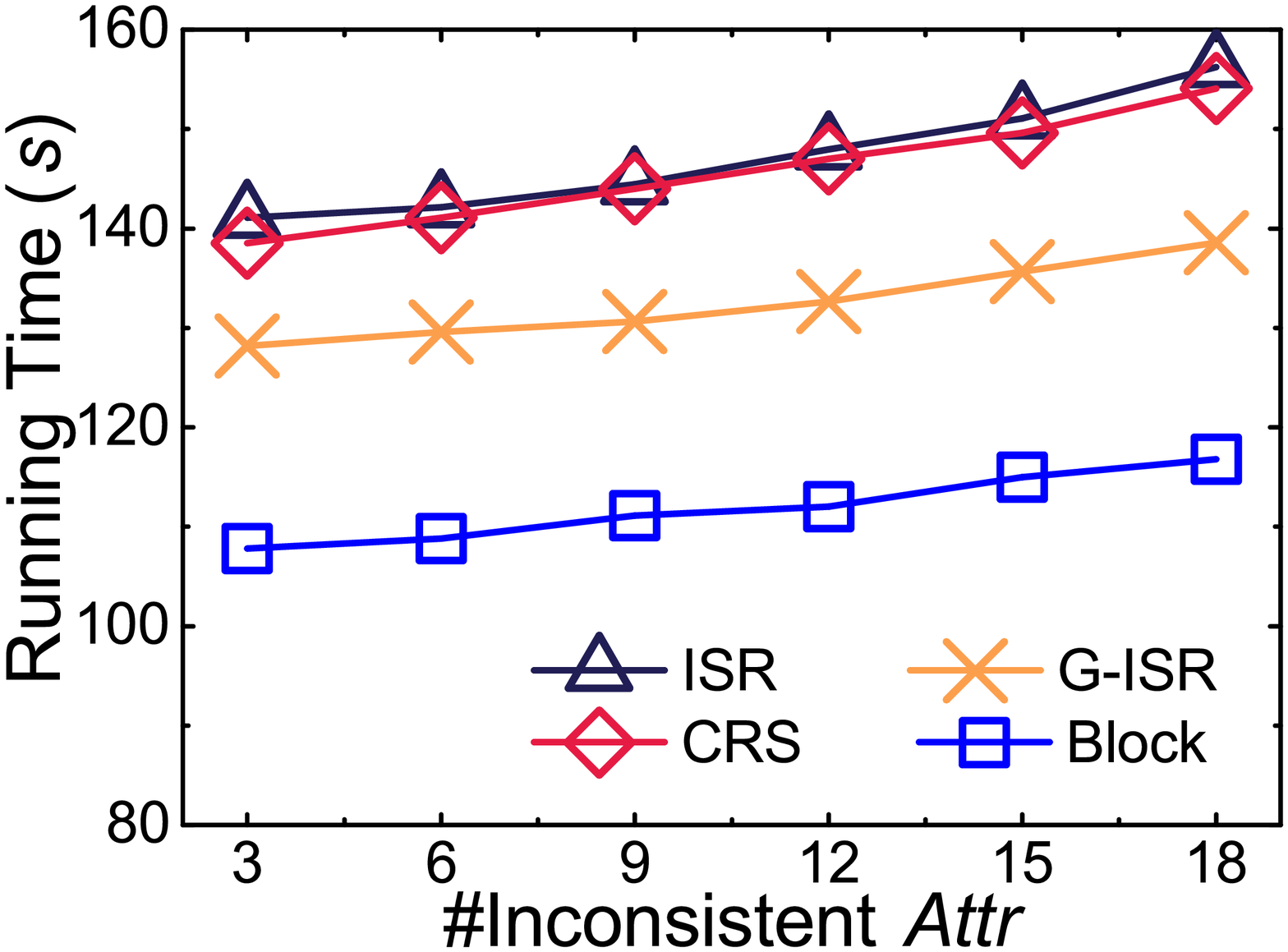}}
\caption{Time costs comparison}
\label{time}
\end{figure}
We report execution time costs of each method with varying data volume in Fig. \ref{time}. Figure \ref{time}(a) shows the total running times on the condition that \#\textsl{Inconsistent Attr} = 12.
Our method $\mathsf{ISR}$ has the highest time cost, which is only a little higher than the time cost of $\mathsf{CRS}$. It is because Algorithm 2 and Algorithm 3 in $\mathsf{ISR}$ spend more time than $\mathsf{CRS}$ to determine the final repair schemas. However, it is easy to see that our repairing schemas determination step does not take much time in each inconsistency solution. It also reflects the Disjoint structure applied in Algorithm 3 does improve the efficiency of determining repair solution. The time costs difference between are certainly to be acceptable, for $\mathsf{ISR}$ achieve better inconsistency repairing  performance than $\mathsf{CRS}$ does. Execution time of $\mathsf{ISR}$ and $\mathsf{CRS}$ increase slower when \#\textsl{Time points} reaches 35\emph{K}, and we are able to finish detection and repairing of inconsistency issues within 5 days' monitoring time series in 2.5 minutes.
\newline \indent $\mathsf{G}$-$\mathsf{ISR}$  have less running times than $\mathsf{ISR}$ and $\mathsf{CRS}$, for the reason that $\mathsf{G}$-$\mathsf{ISR}$ spends less time in graph matching and obtaining candidate repairing schemas with greedy-base method. $\mathsf{Block}$ costs the least time compared with the other methods. It saves much time in the process of bipartite graph construction and matching computation.
\newline \indent  Time costs with varying total number of inconsistent attributes \emph{w.r.t} \#\emph{Time points} = 45\emph{K} in \textsl{FPP}-\textsl{sys} is shown in  Fig. \ref{time}(b).  With the increasing \#\textsl{Inconsistent Attr}, all methods shows
a growing execution time cost. It is because algorithms need more computation (especially in graph matching and determining repair schemas) when repairing complex inconsistency instances among more attributes. Our proposed $\mathsf{ISR}$ has the highest time cost. But $\mathsf{ISR}$ in general can finish inconsistency detection and repairing in 9 attributes on 45\emph{K} time points in 145 seconds. It verifies the effectiveness and efficiency of our method in industrial temporal data cleaning under complex data quality problem scenarios.
\section{Related Work}
\label{newsec2}
We summarize a few works related to our proposed inconsistency issues in time series.
\newline \indent \textbf{Temporal data cleaning}. Data cleaning and repairing is of great importance in data preprocessing, which has been studied extensively. Along with the rise of temporal data mining, temporal data quality issues become serious. Effective cleaning on temporal data is gaining attention according to its valuable temporal information. With the fact that timestamps are often unavailable or imprecise in data application \cite{DBLP:journals/pvldb/ZhangDI10,DBLP:journals/tods/FanGW12}, {the cleaning involves two main problems}: 1) cleaning inconsistent or imprecise timestamps, and 2) repairing anomalous data values and errors. For the former problem, \cite{DBLP:conf/aiia/1993} first proposes a temporal constraints processing framework to address time-related relationship between events.
Song et al. \cite{DBLP:journals/pvldb/SongC016} develops high-quality temporal constraints-based repairing algorithms to solve inconsistent timestamps problems. \cite{DBLP:journals/pvldb/ZhangDI10} proposes a temporal framework to assign possible time interval to each event considering occurrence times of patterns.
For the latter, both statistical-based \cite{DBLP:conf/sigmod/YakoutBE13,Zhang2016Sequential} and constraints-based \cite{DBLP:journals/pvldb/GolabKKSS09,DBLP:conf/sigmod/SongZWY15} cleaning are widely applied in temporal date quality improvement.
\cite{DBLP:journals/pvldb/GolabKKSS09} extends the idea of constraints from dependencies defined on relational database (\emph{e.g}., \textsc{fd}, \textsc{cfd} in \cite{DBLP:series/synthesis/2012Fan}), and proposes sequential dependencies (\textsc{sd}) to describe the semantics of temporal data. Accordingly, speed constraints are developed in sequential data and applied to time series cleaning solutions \cite{DBLP:conf/sigmod/SongZWY15,Zhang2016Sequential}.
\newline \indent \textbf{Anomaly Detection over time series}.
As one common form of temporal data, time series becomes more easily to be collected and further analyzed under data application scenarios. Anomaly detection (see \cite{DBLP:journals/csur/ChandolaBK09} as a survey) is a important step in time series management process \cite{DBLP:journals/tkde/JensenPT17}, which
aims to discover unexpected changes in patterns or data values in time series.
Gupta et al. \cite{DBLP:series/synthesis/2014Gupta} summarizes anomaly detection tasks in kinds of temporal data and provide an overview of detection techniques (\emph{e.g.}, statistical techniques, distance-based approaches, classification-based approaches) in different scenarios. Time series anomaly detection tasks include discovering discrete abnormal data points (outliers) and anomalous (sub)sequences. Autoregression and window moving-average models (\emph{e.g.,} EWMA, ARIMA \cite{DBLP:books/daglib/0070577}) are widely used in outlier points detections \cite{DBLP:journals/tkde/TakeuchiY06}. On the other hand, anomalous subsequences are more challenged to be detected because abnormal behaviors within subsequences are difficult to be distinguished from normal behaviors \cite{DBLP:conf/kdd/ToledanoCBT17}.
Sequence patterns discovery in time series is continuously studied such as \cite{DBLP:conf/vldb/PapadimitriouSF05,DBLP:conf/kdd/Morchen06}.
\cite{DBLP:journals/ml/RebbapragadaPBA09} studies anomalous time series intervals and abnormal subsequences. Further, {high-dimension feature} in time series is taken into account for effectiveness improvement in anomaly detection methods \cite{DBLP:journals/pr/ErfaniRKL16,DBLP:journals/tsmc/LiuLLZ18}.
\newline \indent As the inconsistency problems in industrial temporal data have just been brought to attention in both research and applications of IoT, especially in IIoT scenarios. Technological breakthroughs are still in demand in developing a comprehensive data quality improvement and data cleaning approaches, where inconsistency repairing is a key problem. {In our pervious work \cite{DBLP:journals/pvldb/DingWSLLG19}, we develop a data cleaning systems \emph{Cleanits}, in which we implement reliable data cleaning algorithms about missing value imputation, abnormal subsequence detection and so on.

Our work in this paper develops an integrated data inconsistency repairing method on IoT time series data. The proposed method can also complement existing IoT data cleaning techniques.
\section{Conclusion}
\label{sec6}
We formalize one serious inconsistency problem on multivariate industrial time series data in this paper. We propose an integrated method to detect inconsistent instances and then repair them with correct schemas. The proposed method achieves that: (1) It is effectiveness in IIoT data quality management and data cleaning tasks, (2) Less-negative-cumulative-effect sequence behavior modelling guarantees the reliable of the proposed inconsistency detection process, and (3) Fault-tolerance evaluation on candidate repairing schemas contribute to high-quality repairing on various inconsistency instances. The evaluation results on real-life IIoT data show that the proposed method effectively detects and repairs inconsistency instances in industrial time series within a reasonable time in IoT data monitoring systems.

\ifCLASSOPTIONcompsoc
\else
\fi


\ifCLASSOPTIONcaptionsoff
  \newpage
\fi
\bibliographystyle{IEEEtran}
\bibliography{demo2}

\begin{thebibliography}{10}
\providecommand{\url}[1]{#1}
\csname url@samestyle\endcsname
\providecommand{\newblock}{\relax}
\providecommand{\bibinfo}[2]{#2}
\providecommand{\BIBentrySTDinterwordspacing}{\spaceskip=0pt\relax}
\providecommand{\BIBentryALTinterwordstretchfactor}{4}
\providecommand{\BIBentryALTinterwordspacing}{\spaceskip=\fontdimen2\font plus
\BIBentryALTinterwordstretchfactor\fontdimen3\font minus
  \fontdimen4\font\relax}
\providecommand{\BIBforeignlanguage}[2]{{%
\expandafter\ifx\csname l@#1\endcsname\relax
\typeout{** WARNING: IEEEtran.bst: No hyphenation pattern has been}%
\typeout{** loaded for the language `#1'. Using the pattern for}%
\typeout{** the default language instead.}%
\else
\language=\csname l@#1\endcsname
\fi
#2}}
\providecommand{\BIBdecl}{\relax}
\BIBdecl

\bibitem{DBLP:journals/apin/DingLZ0L16}
J.~Ding, Y.~Liu, L.~Zhang, J.~Wang, and Y.~Liu, ``An anomaly detection approach
  for multiple monitoring data series based on latent correlation probabilistic
  model,'' \emph{Appl. Intell.}, vol.~44, no.~2, pp. 340--361, 2016.

\bibitem{DBLP:journals/pvldb/AbedjanCDFIOPST16}
Z.~Abedjan, X.~Chu, D.~Deng, R.~C. Fernandez, I.~F. Ilyas, M.~Ouzzani,
  P.~Papotti, M.~Stonebraker, and N.~Tang, ``Detecting data errors: Where are
  we and what needs to be done?'' \emph{{PVLDB}}, vol.~9, no.~12, pp.
  993--1004, 2016.

\bibitem{DBLP:conf/kdd/ToledanoCBT17}
M.~Toledano, I.~Cohen, Y.~Ben{-}Simhon, and I.~Tadeski, ``Real-time anomaly
  detection system for time series at scale,'' in \emph{Proceedings of the
  {KDD} Workshop on Anomaly Detection}, 2017, pp. 56--65.

\bibitem{DBLP:series/synthesis/2014Gupta}
M.~Gupta, J.~Gao, C.~C. Aggarwal, and J.~Han, \emph{Outlier Detection for
  Temporal Data}, ser. Synthesis Lectures on Data Mining and Knowledge
  Discovery.\hskip 1em plus 0.5em minus 0.4em\relax Morgan {\&} Claypool
  Publishers, 2014.

\bibitem{DBLP:journals/pvldb/DingWSLLG19}
X.~Ding, H.~Wang, J.~Su, Z.~Li, J.~Li, and H.~Gao, ``Cleanits: {A} data
  cleaning system for industrial time series,'' \emph{{PVLDB}}, vol.~12,
  no.~12, pp. 1786--1789, 2019.

\bibitem{DBLP:journals/access/WangW20}
\BIBentryALTinterwordspacing
X.~Wang and C.~Wang, ``Time series data cleaning: {A} survey,'' \emph{{IEEE}
  Access}, vol.~8, pp. 1866--1881, 2020. [Online]. Available:
  \url{https://doi.org/10.1109/ACCESS.2019.2962152}
\BIBentrySTDinterwordspacing

\bibitem{Lyndon1977Combinatorial}
R.~C. Lyndon and P.~E. Schupp, \emph{Combinatorial group theory}, 1977.

\bibitem{DBLP:phd/ethos/McKay70}
J.~K.~S. McKay, ``Computing with finite groups,'' Ph.D. dissertation,
  University of Edinburgh, {UK}, 1970.

\bibitem{DBLP:books/daglib/0023376}
\BIBentryALTinterwordspacing
T.~H. Cormen, C.~E. Leiserson, R.~L. Rivest, and C.~Stein, \emph{Introduction
  to Algorithms, 3rd Edition}.\hskip 1em plus 0.5em minus 0.4em\relax {MIT}
  Press, 2009. [Online]. Available:
  \url{http://mitpress.mit.edu/books/introduction-algorithms}
\BIBentrySTDinterwordspacing

\bibitem{DBLP:journals/pvldb/ZhangDI10}
H.~Zhang, Y.~Diao, and N.~Immerman, ``Recognizing patterns in streams with
  imprecise timestamps,'' \emph{{PVLDB}}, vol.~3, no.~1, pp. 244--255, 2010.

\bibitem{DBLP:journals/tods/FanGW12}
W.~Fan, F.~Geerts, and J.~Wijsen, ``Determining the currency of data,''
  \emph{{ACM} Trans. Database Syst.}, vol.~37, no.~4, pp. 25:1--25:46, 2012.

\bibitem{DBLP:conf/aiia/1993}
P.~Torasso, Ed., \emph{Advances in Artificial Intelligence, Third Congress of
  the Italian Association for Artificial Intelligence, AI*IA'93, Torino, Italy,
  October 26-28, 1993, Proceedings}, ser. Lecture Notes in Computer Science,
  vol. 728.\hskip 1em plus 0.5em minus 0.4em\relax Springer, 1993.

\bibitem{DBLP:journals/pvldb/SongC016}
S.~Song, Y.~Cao, and J.~Wang, ``Cleaning timestamps with temporal
  constraints,'' \emph{{PVLDB}}, vol.~9, no.~10, pp. 708--719, 2016.

\bibitem{DBLP:conf/sigmod/YakoutBE13}
M.~Yakout, L.~Berti{-}{\'{E}}quille, and A.~K. Elmagarmid, ``Don't be scared:
  use scalable automatic repairing with maximal likelihood and bounded
  changes,'' in \emph{Proceedings of the {ACM} {SIGMOD} International
  Conference on Management of Data, {SIGMOD} 2013, New York, NY, USA, June
  22-27, 2013}, pp. 553--564.

\bibitem{Zhang2016Sequential}
A.~Zhang, S.~Song, and J.~Wang, ``Sequential data cleaning: {A} statistical
  approach,'' in \emph{Proceedings of the International Conference on
  Management of Data, {SIGMOD} Conference}, 2016, pp. 909--924.

\bibitem{DBLP:journals/pvldb/GolabKKSS09}
L.~Golab, H.~J. Karloff, F.~Korn, A.~Saha, and D.~Srivastava, ``Sequential
  dependencies,'' \emph{{PVLDB}}, vol.~2, no.~1, pp. 574--585, 2009.

\bibitem{DBLP:conf/sigmod/SongZWY15}
S.~Song, A.~Zhang, J.~Wang, and P.~S. Yu, ``{SCREEN:} stream data cleaning
  under speed constraints,'' in \emph{Proceedings of the 2015 {ACM} {SIGMOD}
  International Conference on Management of Data, Melbourne, Victoria,
  Australia, May 31 - June 4, 2015}, pp. 827--841.

\bibitem{DBLP:series/synthesis/2012Fan}
W.~Fan and F.~Geerts, \emph{Foundations of Data Quality Management}, ser.
  Synthesis Lectures on Data Management.\hskip 1em plus 0.5em minus 0.4em\relax
  Morgan {\&} Claypool Publishers, 2012.

\bibitem{DBLP:journals/csur/ChandolaBK09}
V.~Chandola, A.~Banerjee, and V.~Kumar, ``Anomaly detection: {A} survey,''
  \emph{{ACM} Comput. Surv.}, vol.~41, no.~3, pp. 15:1--15:58, 2009.

\bibitem{DBLP:journals/tkde/JensenPT17}
S.~K. Jensen, T.~B. Pedersen, and C.~Thomsen, ``Time series management systems:
  {A} survey,'' \emph{{IEEE} Trans. Knowl. Data Eng.}, vol.~29, no.~11, pp.
  2581--2600, 2017.

\bibitem{DBLP:books/daglib/0070577}
W.~W.~S. Wei, \emph{Time series analysis - univariate and multivariate
  methods}.\hskip 1em plus 0.5em minus 0.4em\relax Addison-Wesley, 1989.

\bibitem{DBLP:journals/tkde/TakeuchiY06}
J.~Takeuchi and K.~Yamanishi, ``A unifying framework for detecting outliers and
  change points from time series,'' \emph{{IEEE} Trans. Knowl. Data Eng.},
  vol.~18, no.~4, pp. 482--492, 2006.

\bibitem{DBLP:conf/vldb/PapadimitriouSF05}
S.~Papadimitriou, J.~Sun, and C.~Faloutsos, ``Streaming pattern discovery in
  multiple time-series,'' in \emph{Proceedings of the 31st International
  Conference on Very Large Data Bases, Trondheim, Norway, August 30 - September
  2, 2005}, pp. 697--708.

\bibitem{DBLP:conf/kdd/Morchen06}
F.~M{\"{o}}rchen, ``Algorithms for time series knowledge mining,'' in
  \emph{Proceedings of the Twelfth {ACM} {SIGKDD} International Conference on
  Knowledge Discovery and Data Mining, Philadelphia, PA, USA, August 20-23,
  2006}, 2006, pp. 668--673.

\bibitem{DBLP:journals/ml/RebbapragadaPBA09}
U.~Rebbapragada, P.~Protopapas, C.~E. Brodley, and C.~R. Alcock, ``Finding
  anomalous periodic time series,'' \emph{Machine Learning}, vol.~74, no.~3,
  pp. 281--313, 2009.

\bibitem{DBLP:journals/pr/ErfaniRKL16}
S.~M. Erfani, S.~Rajasegarar, S.~Karunasekera, and C.~Leckie,
  ``High-dimensional and large-scale anomaly detection using a linear one-class
  {SVM} with deep learning,'' \emph{Pattern Recognition}, vol.~58, pp.
  121--134, 2016.

\bibitem{DBLP:journals/tsmc/LiuLLZ18}
H.~Liu, X.~Li, J.~Li, and S.~Zhang, ``Efficient outlier detection for
  high-dimensional data,'' \emph{{IEEE} Trans. Systems, Man, and Cybernetics:
  Systems}, vol.~48, no.~12, pp. 2451--2461, 2018.

\end{thebibliography}

\end{document}